\documentclass[12pt,preprint]{aastex}                    
\shorttitle{\sst Observations of IC348}
\shortauthors{Lada et al.}  
\newcommand{\chisq}{ \ensuremath{ \chi^{2} }}

\newcommand{\av}{\ensuremath{ \mathnormal{A}_{V} }}

\newcommand{\tm}{2MASS}

\newcommand{\sst}{{\sl Spitzer\ }}
\newcommand{\irac}{{\sl IRAC\ }}
\newcommand{\mips}{{\sl MIPS\ }}

\newcommand{\fnu}{\ensuremath{F_{\nu}}}

\newcommand{\Vl}{\ensuremath{{V}_{L}}}
\newcommand{\Rc}{\ensuremath{{R}_{C}}}
\newcommand{\Ic}{\ensuremath{{I}_{C}}}

\newcommand{\Jtm}{\ensuremath{{J}}}
\newcommand{\Htm}{\ensuremath{{H}}}
\newcommand{\Ks}{\ensuremath{{K}_{s}}}

\newcommand{\SIa}{\ensuremath{3.6}}
\newcommand{\SIb}{\ensuremath{4.5}}
\newcommand{\SIc}{\ensuremath{5.8}}
\newcommand{\SId}{\ensuremath{8.0}}
\newcommand{\SMa}{\ensuremath{24}}
\def\msun{M$_\odot$} 
\def\msunsp{M$_\odot$ \ }

\def\cm3{cm$^{-3}$}

\def\13co{$^{13}$CO}

\begin{document} 
\title{ \sst Observations of IC~348:  The Disk Population at
2-3 Million Years.}

\author{Charles J.  Lada, August A.  Muench, K. L. Luhman, 
Lori Allen, Lee Hartmann, Tom Megeath, Philip Myers \& Giovanni Fazio}
\affil{Harvard-Smithsonian Center for Astrophysics} 
\affil{Cambridge, MA 02138}
\email{clada@cfa.harvard.edu, gmuench@cfa.harvard.edu, kluhman@cfa.harvard.edu
lallen@cfa.harvard.edu, tmegeath@cfa.harvard.edu, pmyers@cfa.harvard.edu
lhartmann@cfa.harvard.edu, gfazio@cfa.harvard.edu} 
\author{Kenneth Wood} 
\affil{School of Physics and Astronomy, University of Saint Andrews} 
\affil{Saint Andrews, KY 16 9SS, Scotland, UK} 
\email{kw25@st-andrews.ac.uk} 
\and
\author{James Muzerolle, George Rieke, Nick Siegler \& Erick Young} 
\affil{Steward Observatory, University of Arizona} 
\affil{Tucson, Az 85712}
\email{jamesm@as.arizona.edu, grieke@as.arizona.edu, eyoung@as.arizona.edu} 
%
%
%


\begin{abstract}

We present near and mid-infrared photometry obtained with the \sst {\sl Space
Telescope} of $\sim$ 300 known members of the IC~348 cluster.  We merge this
photometry with existing ground-based optical and near-infrared photometry in
order to construct optical-infrared spectral energy distributions (SEDs) for all
the cluster members and present a complete atlas of these SEDs.  We employ these
observations to both investigate the frequency and nature of the circumstellar
disk population in the cluster. The \sst observations span a wavelength
range between 3.6 and 24 $\mu$m corresponding to disk radii of $\sim$ 0.1 -- 5
AU from the central star. The observations are sufficiently sensitive to
enable the first detailed measurement of the disk frequency for very low mass stars
at the peak of the stellar IMF. Using measurements of infrared excess between 3.6 and 8 
$\mu$m we find the total frequency of disk-bearing
stars in the cluster to be 50 $\pm$ 6\%.  However, only 30 $\pm$ 4 \% of the
member stars are surrounded by optically thick, primordial disks, while the
remaining disk-bearing stars are surrounded by what appear to be optically thin,
anemic disks. Both these values are below previous estimates for this cluster. 
The disk fraction appears to be a function of spectral type and stellar
mass.  The fraction of stars with optically thick disks ranges from 11 $\pm$ 8\%
for stars earlier than K6, to 47 $\pm$ 12\% for K6-M2 stars to 28 $\pm$ 5\% for
M2 - M6 stars.  The disk longevity and thus conditions for planet formation
appear to be most favorable for the K6-M2 stars which are objects of comparable
mass to the sun for the age of this cluster.  
The optically thick disks around later type ($>$ M4) stars
appear to be less flared than the disks around earlier
type stars.  This may indicate a greater degree of dust settling and a more
advanced evolutionary state for the late M disk population.  Finally we find
that the presence of an optically thick dust disk is correlated with gaseous
accretion as measured by the strength of H$\alpha$ emission. A large fraction of
stars classified as CTTS possess robust, optically thick disks and very few of
such stars are found to be diskless.  The majority (64\%) of stars classified as
WTTS are found to be diskless.  However, a significant fraction (12\%) of these
stars are found to be surrounded by thick, primordial disks. These  results
suggest that it is more likely for dust disks to persist in the absence of
active gaseous accretion than for active accretion to persist in the absence of
dusty disks.

\end{abstract} 
%

%

\section{Introduction} 
\label{sec:intro}

Circumstellar disks are of fundamental importance for the physical processes of star
and planet formation.  During protostellar evolution, disks provide the primary
channel for accreting interstellar gas and dust that ultimately make up a star.
Circumstellar disks are also the formation sites of planetary systems such as our own
solar system.  Characterizations of the frequency and lifetimes of such disks is
important for developing an understanding of disk evolution and planet formation.
Circumstellar disks are most readily detected and identified by their infrared
emission.  In particular, the optical-infrared spectral energy distributions (SEDs)
of stars with circumstellar disks are expected to display a characteristic power-law
shape or signature (e.g., Lynden-Bell \& Pringle 1974).  This expectation has been
largely confirmed by observations (e.g., Rucinski 1986; Adams, Lada and Shu 1987,
Beall 1987, etc.)  Moreover, the detailed shape of a disk's SED is also expected to
be sensitive to the structure and mass of the disk (e.g., Adams, Lada \& Shu 1988,
Kenyon and Hartmann 1987, Wood et al 2001, etc).

Because of the difficulty in obtaining the sufficient wavelength coverage necessary
to construct SEDs of disk-bearing stars, most statistical studies of the frequencies
and lifetimes of circumstellar disks have relied on using measurements of infrared
excess at a single wavelength (usually the longest wavelength permitted by ground
based imaging systems, typically either 2.2 microns or 3.4 microns), as a proxy for
disk identification.  Measurements of such near-infrared excesses in embedded
clusters have indicated high disk frequencies for the youngest stars but relatively
short disk lifetimes (e.g., Haisch, Lada \& Lada 2001a).  However, ground-based
measurements based on excess detection at a single wavelength in the near-infrared
are inherently limited; they are useful for detecting the presence of disks but
provide little information about the more detailed nature (mass, structure, etc.)  of
disks (Wood et al.  2001).  Moreover, such measurements are subject to systematic
uncertainties in such assumed parameters as the adopted intrinsic colors of stars and
extinction law assumed and these uncertainties could directly impact the derived
statistics.  The \sst satellite has greatly improved the ability to confidently
identify and study circumstellar disks by making it possible to obtain SEDs for young
stellar objects over an interesting range of wavelength extending well into the
mid-infrared wavelength bands in reasonable amounts of observing time (e.g., Hartmann
et al.  2005; Sicilia-Aguilar et al.  2005). Mid-infrared wavelengths are ideal for
studying disks because these wavelengths are well removed from the peak of the
underlying stellar energy distribution resulting in much greater excesses over the
underlying stellar photosphere.

As part of the \sst GTO program a number of young embedded clusters have been
imaged with both the IRAC and MIPS cameras with the goal of investigating disk
evolution. The first paper resulting from this program examined two clusters
(Tr~37 and NGC~7160) of differing age (4 and 10 Myr respectively) but at the
same distance (900 pc) from the sun (Sicilia-Aguilar et al.  2005).  These
clusters exhibited significantly different disk fractions (48\% and 4\%,
respectively) suggesting significant disk evolution between 4 and 10 Myr in
these systems.  In addition evidence for disk evolution within the clusters was
also suggested by the identification of a small but significant population of
disk bearing stars with optically thin inner disk regions and infrared excesses
only at wavelengths longward of 4.5 $\mu$m.  In this paper we present and
analyze \sst observations of the IC~348 cluster which is the nearest (320 pc),
relatively rich, embedded cluster to the sun and thus a prime target for the
cluster disk evolution program.  Situated at one end of the Perseus molecular
cloud complex, IC~348 is at an intermediate age (2-3 Myrs) for embedded clusters
and only partially embedded in natal gas and dust.  Its proximity has made it an
important laboratory for investigating early stellar evolution and star
formation.  The cluster has been extensively studied and a great deal of effort
has been directed toward determining its membership, primarily through
spectroscopic investigations with the result that most of its members are likely
known and confirmed (e.g., Luhman et al. 1998; 2003; 2005).  The confirmed
membership is relatively large ($\sim$ 300 stars) covering a wide range in mass
from 0.02 to 5 \msunsp with excellent statistical representation.  Because of
the cluster's proximity, \sst observations enable us to accurately measure the
frequency of infrared excess and circumstellar disks around the lowest mass
stars down to and below the hydrogen burning limit.  In this way we are able to
obtain for the first time a robust disk census at the peak of the stellar IMF.

Previous studies of the disk population in this cluster have produced differing
estimates of the disk fraction.  Near-infrared $JHK$ observations produced an estimate
of the disk fraction of $\sim$ 15\% for the population (Lada \& Lada 1995) while
longer wavelength $L$ (3.4 microns) observations yielded a disk fraction of $\sim$
$65\%$ for the 100 brightest members (Haisch, Lada \& Lada 2001b).  Similarly,
H$\alpha$ observations (Herbig 1998) appeared to suggest a disk fraction of $\sim$
50\%.  At the same time searches for millimeter-wave continuum from disks in the
cluster failed to detect a single disk suggesting that as a whole the disk masses in
IC~348 were lower than those that characterize the Taurus cloud complex (Carpenter
2002).  

In this paper we report new measurements on the disk fraction based on \sst
satellite observations at 3.6, 4.5, 5.8, 8.0 and 24.0 microns using the IRAC and
MIPS cameras.  We are able to construct the SEDs of $\sim$ 300 confirmed members of
the cluster and derive improved and more accurate disk frequencies enabling the
investigation of disk evolution in considerable more detail than possible in
previous studies.

\section{Observations and Data Reduction}

As a part of the Guaranteed Time Observations of the IRAC and MIPS instrument
teams, we obtained images of IC~348 at 3.6, 4.5, 5.8, 8.0, and $24~\micron$\
with IRAC and MIPS on the {\it \sst Space Telescope}.  The IRAC data conisted of
two separate observations, a large shallow map and a small deep map on 2004
February 11 and 18 (UT), respectively.  The plate scale and field of view of
IRAC are $1\farcs2$ and $5\farcm2\times5\farcm2$, respectively.  The camera
produces images with FWHM$=1\farcs6$-$1\farcs9$ from 3.6 to $8.0~\micron$\
\citep{faz04}.  The shallow map contained a $6\times7$ mosaic of pointings
separated by $280\arcsec$ and aligned with the array axes.  At each cell in the
map, images were obtained in the 12~s high dynamic range mode, which provided
one 0.4~s exposure and one 10.4~s exposure.  The map was performed twice with
offsets of several arcseconds between the two iterations.  The resulting shallow
maps had centers of $\alpha=3^{\rm h}43^{\rm m}24^{\rm s}$,
$\delta=32\arcdeg07\arcmin03\arcsec$ (J2000) for 3.6 and $5.8~\micron$\ and
$\alpha=3^{\rm h}44^{\rm m}18^{\rm s}$, $\delta=32\arcdeg13\arcmin47\arcsec$
(J2000) for 4.5 and $8.0~\micron$, dimensions of $33\arcmin\times29\arcmin$, and
position angles of $170\arcdeg$ for the long axes.  The deep IRAC map contained
a $3\times3$ mosaic of pointings separated by $290\arcsec$.  At each cell in the
map, two 100~s images were obtained at each of eight dithered positions,
resulting in a total exposure time of 1600~s for most of the map.  The resulting
deep maps had centers of $\alpha=3^{\rm h}44^{\rm m}44^{\rm s}$,
$\delta=32\arcdeg05\arcmin50\arcsec$ (J2000) for 3.6 and $5.8~\micron$\ and
$\alpha=3^{\rm h}44^{\rm m}37^{\rm s}$, $\delta=32\arcdeg12\arcmin27\arcsec$
(J2000) for 4.5 and $8.0~\micron$, dimensions of $15\arcmin\times15\arcmin$, and
position angles of $79\arcdeg$.

For each of the shallow and deep IRAC maps, the images from the \sst Science
Center pipeline (version S10.5.0) were combined into one mosaic at each of the four
bands using custom IDL software developed by Robert Gutermuth.  Point sources in
these data were identified with the task DAOFIND under the IRAF package APPHOT.
Aperture photometry for these sources then was extracted with the task PHOT using a
radius of two pixels ($2\farcs4$) for the aperture and inner and outer radii of two
and three pixels for the sky annulus.  We selected these relatively small apertures
and sky annuli to provide the best possible subtraction of background emission from
the parent cloud of IC~348, which is bright and spatially variable at mid-IR
wavelengths.  

For an aperture radius of 10 pixels and a sky annulus extending from 10 to 20 pixels, we
adopted zero point magnitudes ($ZP$) of 19.670, 18.921, 16.855, and 17.394 in the 3.6, 4.5,
5.8 and 8~$\mu$m bands, where $ M = -2.5 \log (DN/sec) + ZP$ \citep{rea05}.  We then
applied aperture corrections of 0.274, 0.298, 0.543, and 0.769~mag to the photometry.  The
photometry from the shallow and deep images were merged into one catalog.  We used sources
found in both catalogs to characterize our photometric uncertainties.  A total of
764, 817, 308 and 223 overlapping sources were found in bands 1-4, respectively.  Of these
$84$, $79$, $90$ and $83\%$ had observed 1 sigma scatter less than 0.1 magnitudes.  Only
$1-2\%$ of the overlap sources had 1 sigma variations which were greater than the 3 sigma
predicted photometric errors calculated from the aperture photometry and which includes sky
background, read and shot nosie.  These estimates of the photometric accuracy are
particularly relevant for known cluster members, which fall predominately within this
overlap region.  In this combined data set, one bright star, HD~23180 ($V=3.9$), lacks
unsaturated measurements at 3.6 and $4.5\;\micron$.

Observations of IC 348 using the Multiband Imaging Photometer for {\it Spitzer}
(MIPS; Rieke et al.  2004) were taken in February 2004.  The cluster was mapped
using scan mode with 12 0.5-degree scan legs and half-array cross-scan overlap,
resulting in a total map size of about 0.5 x 1.5 degrees including overscan and
overlapping completely the IRAC observations.  The total effective exposure time
per point at $24 \micron$\ is about 80 seconds.  The MIPS images were processed using
the instrument team Data Analysis Tool, which calibrates the data and applies a
distortion correction to each individual exposure before combining into a final
mosaic (Gordon et al.  2005).  Point source photometry was performed on the 24
$\mu$m mosaic using PSF fitting with {\it daophot}.  Except in a few cases where
faint sources lay near bright extended background emission, the measurement errors
are dominated by the 10\% uncertainty in the absolute flux calibration.  We do not
consider data from the other two MIPS channels in this contribution.

\section{Results}

\subsection{Membership Sample}

In this paper we consider only the observations of $\sim\ 300$ cluster stars with confirmed
membership by Luhman et al (2003).  These cluster members were initially selected from a
complete magnitude-limited sample of stars whose optical-infrared colors placed them above
the main sequence for the distance of this cluster.  Spectra and near-infrared photometry
exists for a significant fraction of these candidate members.  Membership for these stars
was assessed on the basis of the following factors:  known proper-motions, the presence of
Li absorption, spectroscopic indication of low surface gravity, photometric evidence of
extinction, presence of emission lines and/or infrared excess.  The membership sample is
thus not biased toward disk bearing stars since the selection was not limited to stars with
observed signatures of disks such as emission-line or infrared-excess stars.  The sample is
known to be complete within a 16 by 14 arc minute region centered on the cluster for stars
with masses greater than 0.03 solar masses and for visual extinctions less than 4
magnitudes (Luhman et al.  2003).  This area encompasses 75\% of the confirmed members
examined in this paper.  In Table \ref{tab:catalog} we present a catalog of the IRAC and
MIPS photometry for all sources.  We combined our \sst\ \irac\ and \mips\ data with
ground-based near-infrared and optical observations from the literature to construct
spectral energy distributions (SEDs) for all the sources in the catalog.  We present an
atlas of these SEDs for the members of IC~348 in Appendix \ref{app:atlas}.

\subsection{Spectral Energy Distributions}

\begin{figure}
    \centering
   \begin{minipage}[c]{\textwidth}
     \centering \includegraphics[angle=0,width=0.8\textwidth]{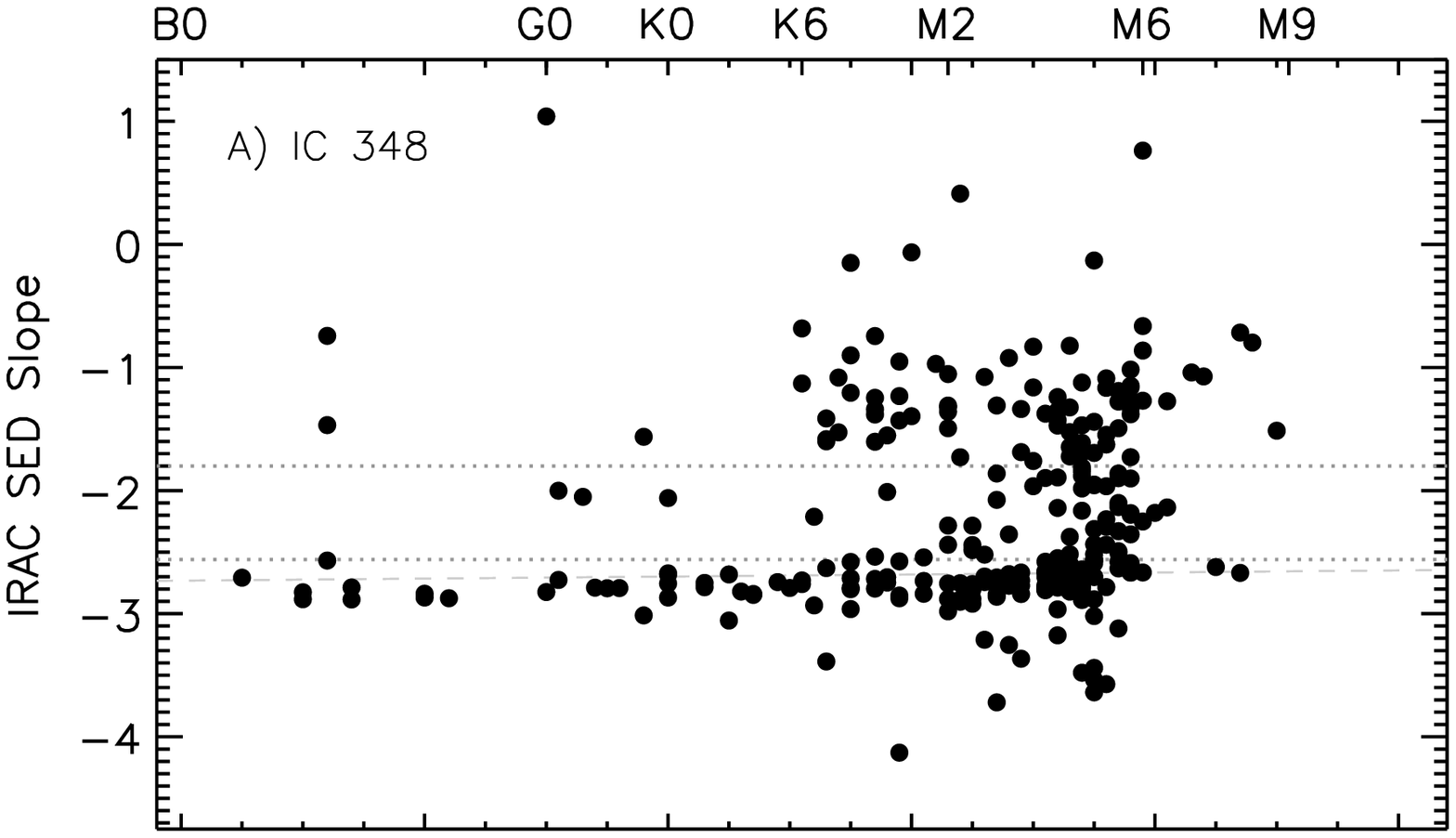}
    \end{minipage}%
   \hfill%
    \begin{minipage}[c]{\textwidth}
     \centering \includegraphics[angle=0,width=0.8\textwidth]{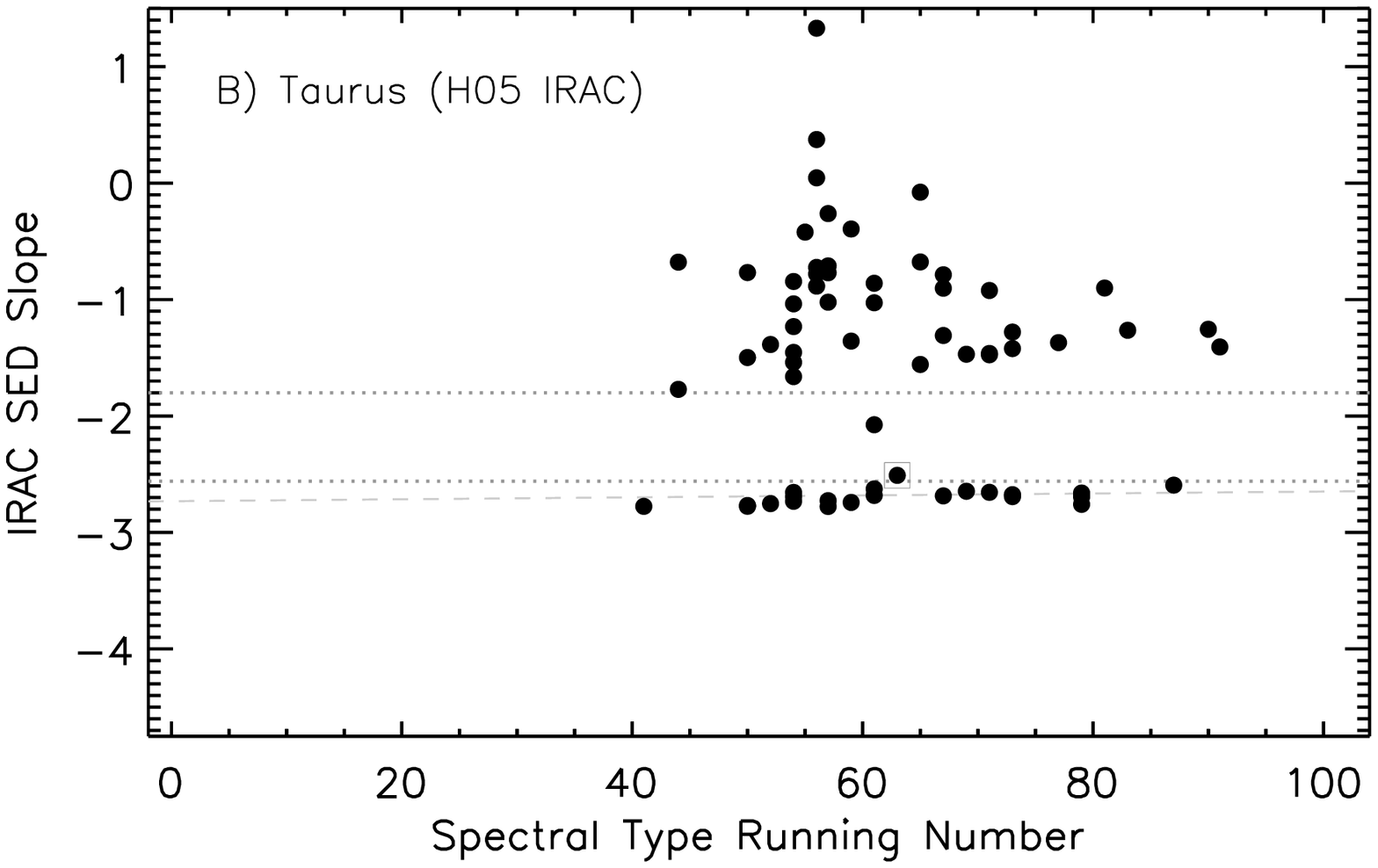}
    \end{minipage}%
\caption{The dereddened power-law slopes of the IRAC SEDs for young stars
plotted  as a function of spectral type. A) Members of the IC~348 cluster;(B)
the corresponding  diagram for young stars in Taurus.  Two prominent bands are
observed in each diagram corresponding to stars with optically thick
circumstellar disks and diskless stars.  Stars whose slopes fall between the two
bands are anemic disks or transition objects; for example, the Taurus M1.5 star
CoKu Tau-4, which has has been proposed to have a inner hole of radii 10 AU
\citep{2005ApJ...621..461D}, can be identified as such a transition object 
in panel (B), having an $\alpha_{IRAC}=-2.51$ (open square). Data for Taurus
stars taken from Hartmann et al. 2005 (H05).}
\label{sedslopes} 

\end{figure}

The examination of the optical-infrared SED of a target star provides the most robust
method for the identification of the presence of a circumstellar disk.  The shape of the
SED can distinguish a disk origin for observed infrared excesses from other possible causes
and moreover is sensitive to such parameters as disk structure, orientation and if extended
to millimeter-wavelengths, the size distribution of emitting particles.  A particularly
useful measure of the shape of an SED is its slope (e.g., Lada 1987).  Here we use the $3.6
-- 8.0\; \micron$\ slope of the IRAC SED for each of the 263 sources in our catalog/atlas
that was detected in all four IRAC bands to identify stars with disks and obtain a census
of the circumstellar disk population in the cluster.  We obtain our metric for the SED
slope, $\alpha$, from a simple power-law, least-squares fit to the four IRAC bands for each
star.  The results of this procedure are shown in Figure \ref{sedslopes} where we plot the
slopes we derived for the stars in IC~348 as a function of spectral type.  For reference we
also derive the same diagram for stars in Taurus with published IRAC data (Hartmann et al.
2005).

In both clouds the sources primarily cluster around two distinct bands, one centered
on an $\alpha$ $\approx$ -1.0 and the other centered on an $\alpha$ $\approx$ -2.7
corresponding to disk-bearing and diskless stars, respectively.  Examination of the
SED slopes in Taurus shows a distinct separation between objects within the two
bands.  This separation between diskless stars and disk-bearing stars has been
noticed previously and has been interpreted to indicate a rapid transition between
the two phases of evolution (Kenyon \& Hartmann 1995; Wood et al.  2002).  In the SED
atlas in Appendix \ref{app:atlas} we compare individual source optical-infrared SEDs
to two disk models calculated from a Monte Carlo radiative transfer code (Bjorkman \&
Wood, 2001; Wood et al.  2002; Walker et al.  2004).  The model disks extend from the
dust destruction radius to 200 AU and are purely reprocessing disks with no
accretion.  For both models the disk mass is 3\% of the stellar mass and the radial
dependence of the disk surface density is $\Sigma (r) \sim r^{-1}$.  The dust size
distribution used in the models is described in Wood et al. (2002).  One model is
flared consistent with the requirements of vertical hydrostatic equilibrium while the
other has a scale height reduced by a factor of 3 to approximate a disk which is
spatially flatter as would be expected for a disk that has experienced significant
mid-plane settling of its emitting dust.  From a comparison of the observed SEDs, SED
slopes and models we find that sources with disks that match the models (flared or
flat) are characterized by $\alpha$ $>$ -1.8.  From an examination of the SED slopes
of apparently diskless stars in Taurus, we find that diskless stars (i.e., stellar
photospheres) can be characterized by $\alpha$ $<$ -2.56.  We note that the value of
$\alpha_{IRAC}$ for stellar photospheres can vary with spectral type.  Indeed, we
have examined the dependence of $\alpha_{IRAC}$ on spectral type for dereddened
diskless stars in Taurus and found that it can be well approximated by a linear
function which is displayed in panels (A) and (B) of Figure \ref{sedslopes}.  The
variation in $\alpha$ is small across the range of spectral types under
consideration.  Therefore for simplicity, we adopt a single photospheric value of
$\alpha_{IRAC, photosphere}\;=\;-2.56$, which corresponds to the predicted slope for
an M0 star (-2.66) plus a buffer of 0.1 to account for the typical precision of these
SED fits (See, for example, Table \ref{tab:alphafits}).  Sources characterized by
-2.56 $<$ $\alpha$ $<$ -1.80 we term anemic disks and are possibly transition
objects, such as heavily depleted, optically thin disks, disks with inner holes, etc.
(D'Alessio et al. 2005).

In Figure \ref{med_seds} we plot the median SEDs for each of the three SED types
(thick disk, anemic disk, diskless photosphere) for the different spectral classes
(See Appendix \ref{app:median} for the derivation of IC~348 median SEDs).  We note
that anemic disks display very small excesses primarily only at the longest
wavelengths and even where detected they typically have fluxes below that expected
for an optically thick disk.  Comparison with the median SED of Taurus disk-bearing
stars with spectral types between K2-M0 indicates that for the stars in this
spectral type range the circumstellar disks in IC 348 and Taurus are very similar in
nature.  Although for M0-M2 the median IC348 SED falls below the Taurus SED
suggesting perhaps that some disk evolution or dust settling may have occured in
IC~348, similar to the situation in the 4 Myr cluster TR~37 (Sicilia-Aguilar et al.
2005).

\begin{figure}
\plotone{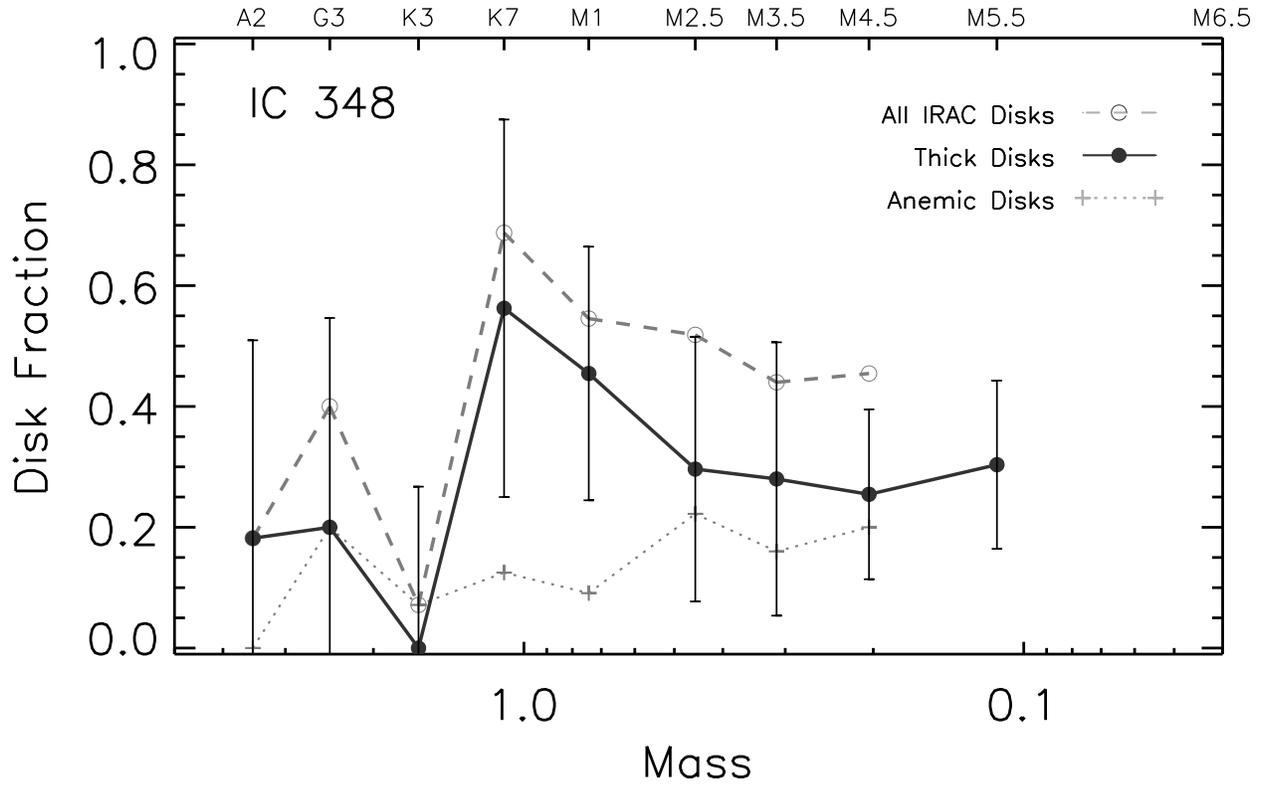}
\caption{Disk frequency as a function of mass and spectral type as derived from
analysis of the slopes of the IRAC SEDs for those cluster members detected in
all four IRAC bands. See text.  The solid line plots the disk fraction for stars
with substantial optically thick disks, the lower dotted line plots the fraction
of stars with anemic (possibly optically thin) disks.  The dashed line on the
top plots the total disk fraction (optically thick $+$ anemic) for the cluster.
The error bars represent simple poisson statistical uncertainties. }

\label{diskfreq_sed} 
\end{figure}

We compute the disk fractions as a function of spectral type and mass and plot the result
in Figure \ref{diskfreq_sed}.  Here we display the fraction of sources for which $\alpha$
$>$ -1.80 (optically thick disks), the fraction of sources with -1.8 $>$ $\alpha$ $>$ -2.65
(anemic or thin disks) and the total disk fraction as a function of spectral type and
stellar mass.  The stellar mass is derived by converting the spectral types to effective
temperatures using a subgiant temperature scale (Luhman 1999) and the effective
temperatures to masses using the Baraffe et al.  (1998) pre- main sequence evolutionary
model calculations appropriate for a 2-3 Myr cluster.  This is a satisfactory conversion
for our purposes since at these ages nearly all these stars are on convective tracks in the
HR diagram.  The disk fraction appears to be a function of spectral type (mass).  The disk
fraction is lowest for stars with spectral types earlier than K6 and highest for stars with
spectral types between K6-M2, that is, for stars with masses around 1 solar mass.  For
later spectral types there appears to be a definite decline in the disk fraction.  Given
the level of uncertainty on the individual points, the significance of this drop is
difficult to assess.  However the fact that the four contiguous lowest mass bins are
systematically lower than the preceeding two bins suggests that the fall off is real.  To
better assess the variation of disk fraction with spectral type we can calculate the disk
fractions over wider intervals of spectral type and mass.  More specifically, for all stars
earlier than K6 we find disk fractions of 11 $\pm$ 8\% and 8 $\pm$ 8\% for optically thick
disks and anemic disks, respectively.  For stars between K6-M2 we find the corresponding
disk fractions to be 47 $\pm$ 12\% and 9 $\pm$ 5\% and for stars between M2-M6 (which make
up the last four bins of Figure \ref{diskfreq_sed}) we find the corresponding disk
fractions to be 28 $\pm$ 5\% and 26 $\pm$ 5\%.  The decline of the optically disk fraction
for the later spectral type/lowest mass stars is apparently significant.  The fraction of
all cluster members detected in all four IRAC bands with optically thick disks is found to
be 30 $\pm$ 4\%, while the total disk fraction (optically thick $+$ anemic disks) for this
population of sources is 50 $\pm$ 6\%.  We note here that our disk census applies to the
entire cluster membership.  If we restrict our sample to only those cluster members located
within the inner 16 by 14 arc min region of the cluster known to be complete for $m_*$
$>$ 0.03 \msun (Luhman et al.  2003), we derive nearly identical disk
fractions as those computed above for the larger sample but with somewhat
larger uncertainties.

\subsection{\sst Color-Color Diagrams}

\begin{figure}
    \centering
   \begin{minipage}[c]{\textwidth}
     \centering \includegraphics[angle=0,width=\textwidth]{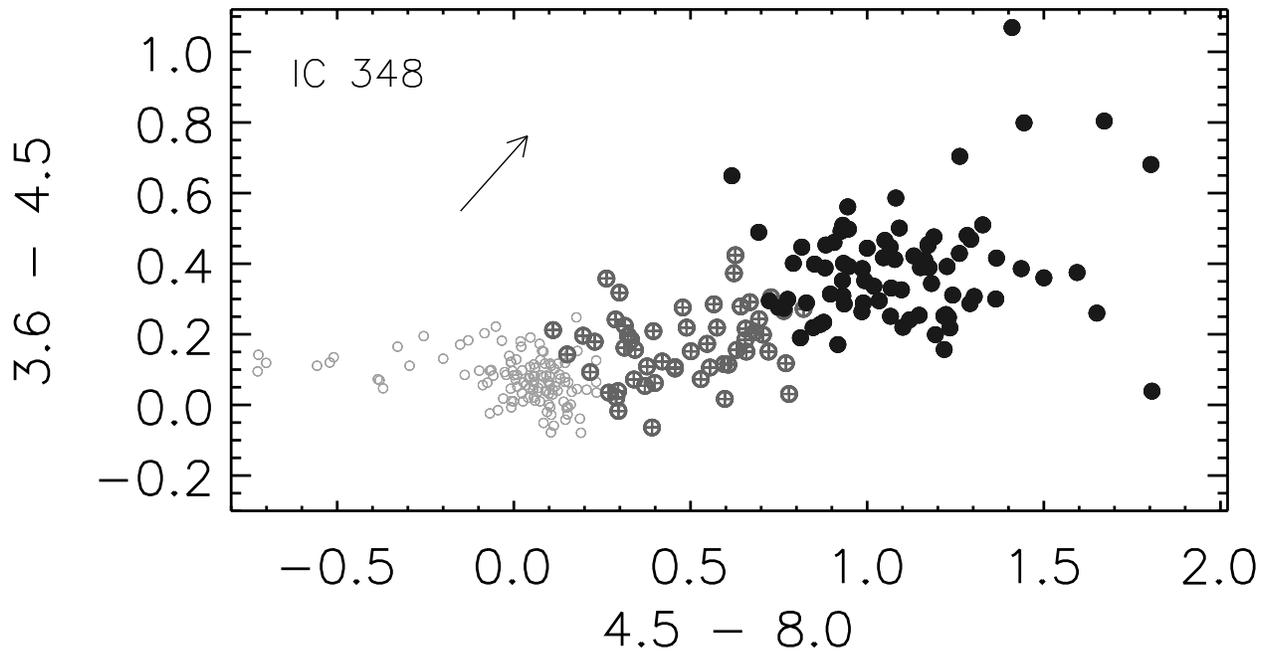}
    \end{minipage}%
\caption{IRAC color-color diagram for IC~348 cluster members. The different
symbols correspond to diskless stars (small open circles), anemic or thin
circumstellar disks (circled crosses) and optically thick disks (filled circles)
as defined by their IRAC SED slope. A reddening vector of length
$\av=20$~magnitude is shown.}

\label{irac_cc}
\end{figure}

In Figure \ref{irac_cc} we plot the IRAC color-color diagram for the cluster.  We
have labeled the points in the figure with different symbols which correspond to
stars whose IRAC SED slopes indicated whether they were diskless or surrounded by
thick or anemic disks respectively.  There is a continuous sequence in color from
diskless stars to stars with progressively large excesses and increasingly
substantial disks.  The stars are clearly spatially segregated according to the
nature of their surrounding disks, Unlike previous IRAC only studies
of embedded clusters we find the gap between disk-bearing and diskless stars
to be filled by stars with anemic disks.

\begin{figure}
\plotone{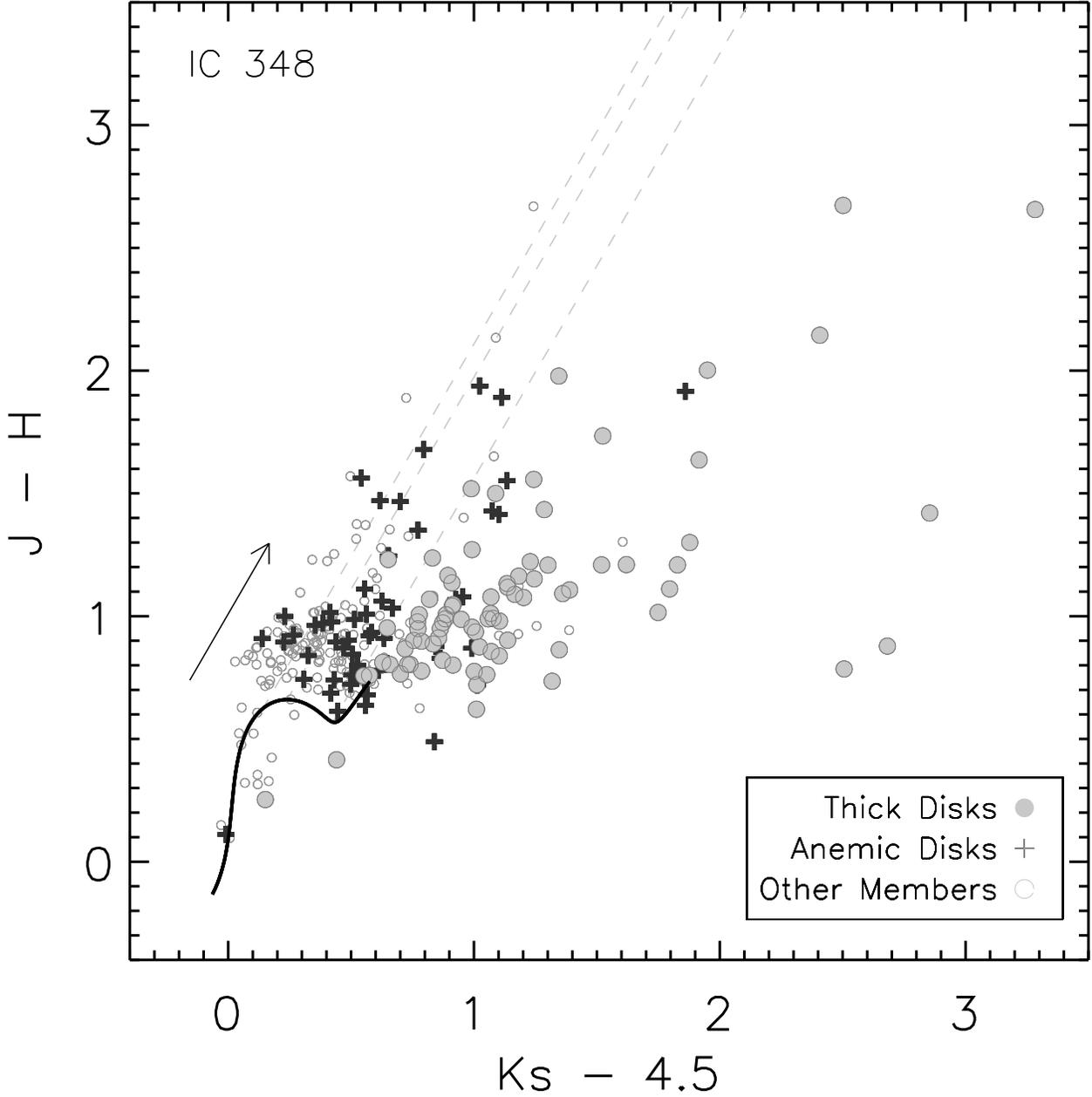}
\caption{The $J(1.25\micron)$, $H(1.65\micron)$, $K(2.2\micron)$, 
$4.5\micron$~color-color diagram for IC~348 cluster members.  Sources are
segregated based upon their \irac\ SED slope; open circles correspond either to
members without IRAC disk excess or to members lacking detections $>5\micron$.  
Also plotted are the locus of colors for main sequence stars (solid line) and 
the reddening boundaries (dashed lines) for three different dwarf spectral
classes: M0, M3 and  M6.  These were derived by merging the \sst data from
the Pleiades main-sequence stars (Stauffer 2005, private communication) and
field M dwarfs (Patten 2005, private communication).
A reddening vector of length $\av=5$ is also shown.}

\label{jhk4_cc_all}
\end{figure}

\begin{figure}
\plotone{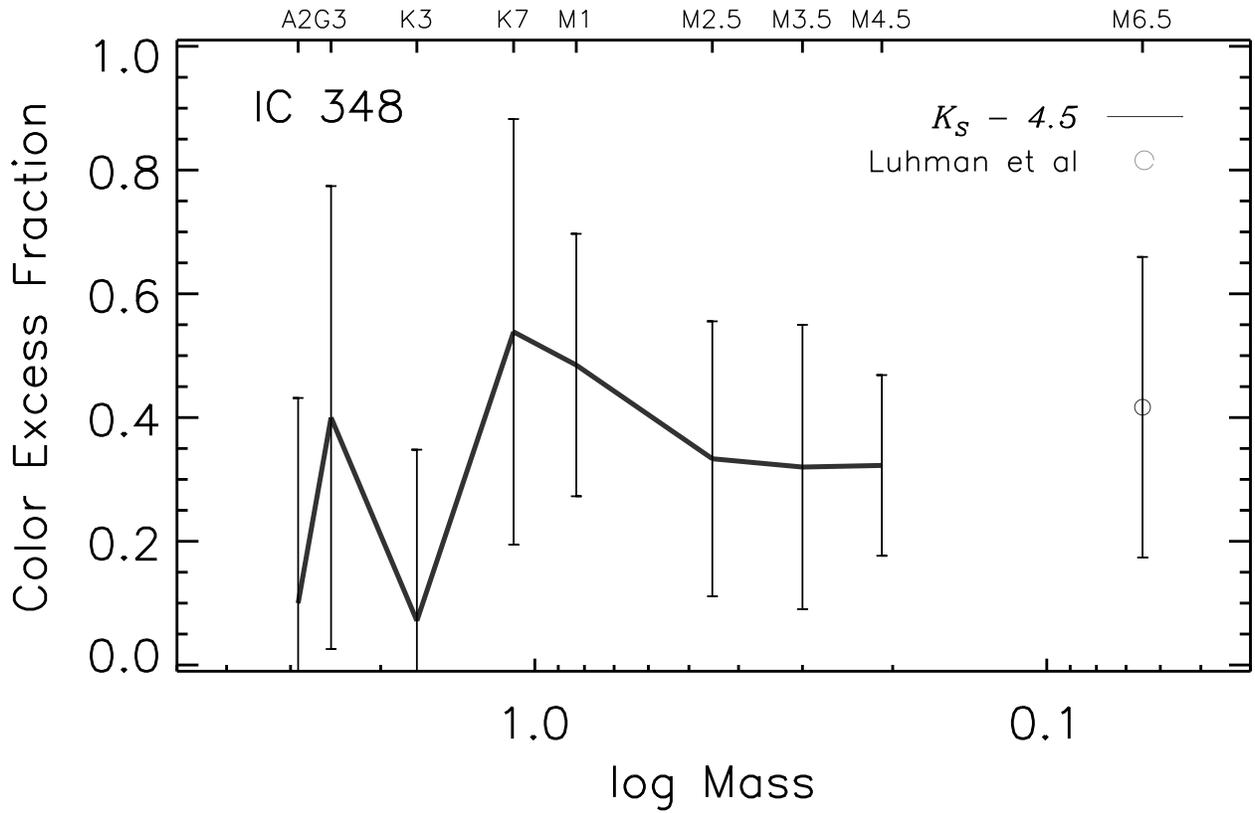}
\caption{The disk frequency for IC~348 as a function of spectral type and mass
derived from analysis of the $JHK\;\;4.5\micron$~color-color diagram (Figure
\ref{jhk4_cc_all}). The brown dwarf disk excess fraction from \citet{luh05c}
with similarly derived error bars is plotted for comparison.}
\label{diskfreq_cc} 
\end{figure}

Not all cluster members were detected in all IRAC bands, however, essentially all the
members were detected at 4.5 $\mu$m.  Consequently, to obtain a more complete disk census
for the entire cluster membership, we need to rely on analysis of shorter wavelength data
and employ color-color diagrams.  We have used the data in Table 1 to construct color-color
diagrams combining $JHK$ photometry with $4.5\;\micron$\ IRAC photometry for the IC~348
cluster members in order to identify sources with infared excesses and compile a more
complete excess/disk census for the cluster.  Infrared excesses are derived in color-color
diagrams from comparison of the observed colors with the intrinsic colors of normal
diskless stars.  We determined the intrinsic colors for the \sst bands by using \sst
observations of Pleiades main-sequence stars (Stauffer 2005, private communication) and
field M dwarfs (Patten 2005, private communication).  In Figure \ref{jhk4_cc_all} we
present the $J(1.25 \micron)$, $H(1.65 \micron)$, $K(2.2 \micron)$, B2~$(4.5 \micron)$
diagram for the cluster.  Also plotted is the locus of main sequence dwarf stars and the
reddening boundaries corresponding to three different spectral types (M0, M3, and M6 ).
These boundaries are trajectories that parallel the reddening vector for stars of the
differing spectral types.  Although Figure \ref{jhk4_cc_all} shows the reddening boundaries
for three spectral types, we have calculated the total disk fraction for the cluster as a
function of spectral type by using the appropriate reddening boundary for eight spectral
type bins from A stars to late M stars and counting the number of stars to the right of
that boundary.  The disk fraction determined in this manner is plotted in Figure
\ref{diskfreq_cc}.  The result for the entire population is very similar to that derived
above using SED slopes for the somewhat more restricted sample of sources detected in all
four IRAS bands.  This Figure confirms the decline in disk fraction with spectral type
below a solar mass.  The fact that the disk fraction derived in this way is lower than the
total disk fraction from figure \ref{diskfreq_sed} indicates that most anemic disks do not
show excess at $4.5\; \micron$.  We did not calculate the disk fraction for stars later than
M6 since this was done in an earlier contribution by Luhman et al.  (2005) who employed
IRAC only color-color diagrams to derive the disk fraction for the brown dwarf population
in IC 348 from the \sst observations.  They found the disk fraction for stars later than M6
to be 42 $\pm$ 13\%.  This is higher than the disk fraction derived for the earlier type M
stars and the cluster as a whole and may indicate a slight increase in the disk fraction
below the hydrogen burning limit.

In figure \ref{jhk4_cc_all} the stellar colors are plotted as different symbols
with the filled circles representing stars found to have optically thick disks
from analysis of their SEDs while the plus signs correspond to stars identified
as possessing anemic disks. The diagram suggests that a reasonably accurate
census of the thick disk fraction can be measured by adopting the M6 
reddening boundary without consideration of the individual spectral types of the 
target stars. Thus if spectral types are not available using the $JHK 4.5\mu$m 
color-color diagram with the M6 reddening boundary can produce a good estimate
of the primordial disk fraction for a young stellar population.

The disk fractions we derive for the IC~348 cluster in Figure \ref{diskfreq_cc} are
significantly lower than that ($\sim$ 50\%) derived by Haisch et al.  (2001b) using
ground-based $L$ band observations in a similar analysis.  The difference is
attributable to two factors.  First, Haisch et al.  examined a magnitude limited ($L
< 12$ mag.)  sample of 107 stars, about a third the size of the sample considered
here.  The \sst sample is dominated by late M stars which have a lower disk fraction
than the K and early M stars that dominated the magnitude-limited Haisch et al.
sample.  Second, and more significant, is that there appears to be a difference in
the observed 3 $\mu$m colors of the stars in the two samples.  If we examine the 95
sources in common between our two studies we derive a disk fraction of 36\% from an
analysis of the $JHK$ \irac\ 3.6 $\mu$m color-color diagram using the same boundary for
the reddening band as adopted by Haisch et al.  This difference in derived disk
fraction appears to be largely the result of the fact that the $K - 3.6 \mu$m colors
of the stars in the present study appear to be bluer than the $K - L$ colors of Haisch
et al.  The origin of this difference is not clear but may be due to systematic
calibration uncertainties the 3.6 $\mu$m observations. However based on analysis
of the stellar SEDs, it appears that the IRAC data produce the more a reliable 
measure of the disk/excess fraction.

\begin{figure}
\plotone{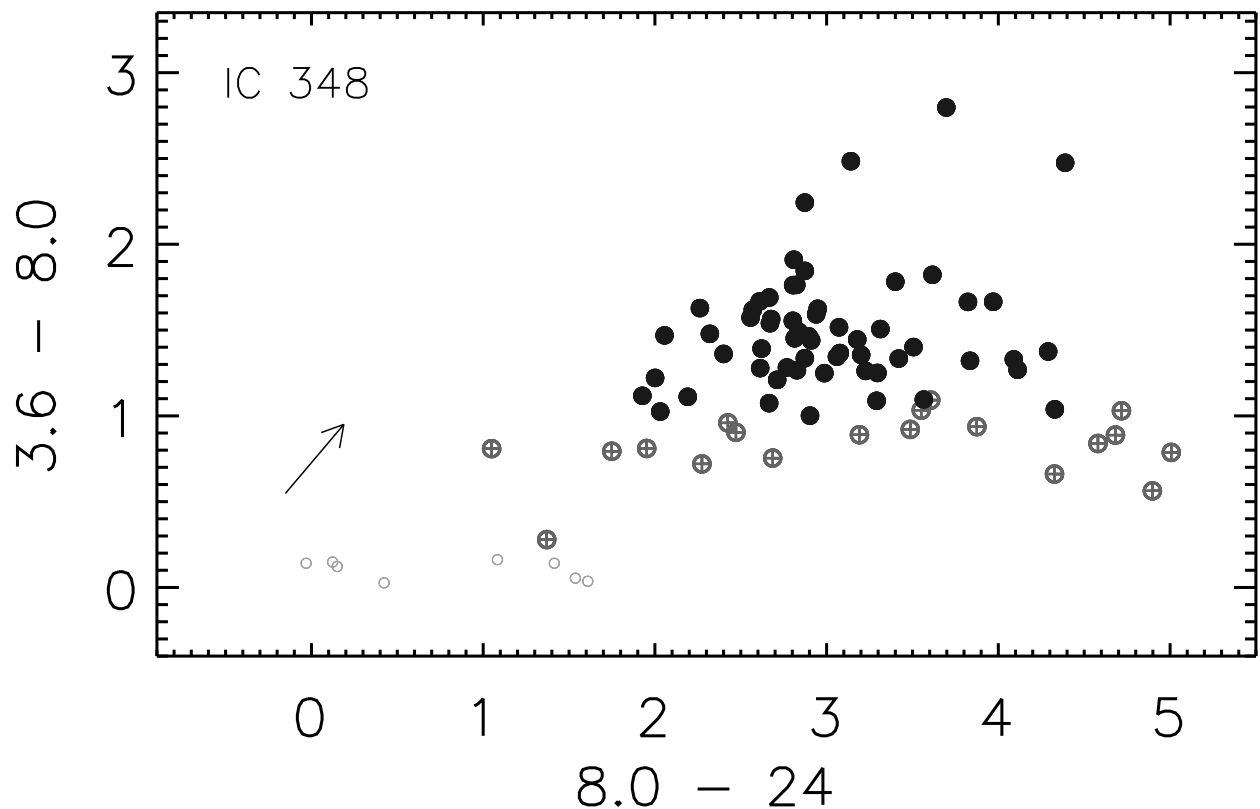}
\caption{IRAC-MIPS color-color diagram for IC~348 sources detected at 
$24\;\micron$. Small open circles indicate diskless stars, circled crosses
indicate anemic disks and solid filled circles denote stars with thick disks.
This figure shows that four early type stars that appear naked or diskless in
IRAC colors show significant MIPS $24\micron$~excess possibly indicating the
presence of a debris disk. Stars with anemic disks form a narrow horizontal band
marking the lower bound of the extent of thick disk sources. Six anemic disks
with the largest $24 \micron$ excess are likely disks with holes or
significantly depleted inner regions.  A reddening vector of length
$\av=20$~magnitude is shown.}
\label{irac_mips}
\end{figure}

In Figure \ref{irac_mips} we plot an IRAC-MIPS color-color diagram for those sources in the
cluster detected at $24\; \micron$.  Different symbols represent the different disk types
(thick and anemic) as well as diskless stars as determined from IRAC colors.  Sources with
thin or anemic disks form a narrow band with little dispersion in 3.6-$8.0\; \micron$ color
but a wide extent in $8.0-24.0\; \micron\ $ color.  This band forms the lower boundary to
the extent of thick disk sources in this color-color plane.  The distribution of anemic
disk sources appears to be less concentrated in the $8-24$ \micron\ color than that of the
thick disk sources.  For the anemic disks the contrast between inner and outer disk
emission increases as the $8-24\; \micron$\ color increases.  The six anemic sources with
the largest $24\; \micron$\ excesses represent objects with the largest contrast between
inner and outer disk emission and may be disks with significant inner holes.  The majority
of anemic disk sources do not appear to possess such significant inner holes.

We also note the presence in Figure \ref{irac_mips} of two small groupings of stars whose
$3.6-8.0\;\micron\ $ color is close to zero magnitude.  Eight out of nine of these stars
show no indication of infrared excess in any of the IRAC bands, while one (source \# 8, see
Table \ref{tab:catalog}) is identified as an anemic disk by its SED slope.  Three of these
sources have $8 - 24\; \mu$m colors close to zero magnitude, characteristic of diskless
photospheres out to 24 $\mu$m wavelength.  Two are G stars (\#s 11 \& 22) and one is a K
star (\# 10363).  A fourth source, an A star (\# 3), also exhibits a near photospheric $8 -
24\; \mu$m color but appears to have a slight excess in the 24$\mu$m band.  A second grouping
of five stars is clustered around an $8 - 24\; \mu$m color of $\sim$ 1.4 magnitudes.  Four of
these stars appear to be diskless in the IRAC bands but clearly exhibit excess at 24 
$\mu$m.  Two of the stars are early type A stars (\#s 7 \& 25) while another is an F star
(\# 30) and one is a G star (\#20).  These stars could either be extreme examples of anemic
(transiton) disks or very young debris disk systems.  To distinguish between these two
possible interpretations requires knowledge of whether the circumstellar material producing
the 24 $\mu$m excess is remnant primordial disk material or newly generated dust produced by
collisions between planetesimals and this is very difficult to ascertain with the exisitng
observations for such a young cluster.

\section{Discussion}

\subsection{Disk Evolution}

We have obtained a detailed census of circumstellar disks among the members of the IC~348
cluster using observations from the \sst Satellite coupled with extensive ground-based data
from the literature.  The total disk fraction for the cluster membership is found to be 50
$\pm$ 6\% but only 30 $\pm$ 4\% of the members bear optically thick primordial disks.  Both
these fractions are lower than previously estimated from ground-based $L$ band
observations, although they are consistent with a normal extrapolation of the measured
ground-based K $(2.2\micron)$ band excess fraction (Lada \& Lada 1995).  Our observations
provide evidence once again for rapid disk evolution in clusters.  If all stars in this
cluster started out with circumstellar disks, then our data indicate that between roughly
$60-70 \%$ of the cluster members have lost their primordial disks in only 2-3 Myrs.
However we also find that the disk fraction for this cluster is also a function of spectral
type and thus stellar mass.  The early type stars exhibit a very low disk fraction
consistent with earlier findings (Haisch et al.  2001b) and suggesting that $\sim$ 90 \% of
stars earlier than K6 have lost their disks in only 2-3 Myrs.  The disk fraction peaks for
late K and early M stars, which are solar mass objects at the age of the cluster and
roughly 50\% of these solar analogs have managed to retain their original disks over the
lifetime of the cluster.  Surprisingly the disk fraction drops for later spectral type
stars that dominate the cluster membership.  Only about 30\% of these stars have retained
optically thick disks, suggesting perhaps that their rate of evolution is between that of
the early type stars and solar type stars.

An expected consequence of disk evolution is a change in disk structure due to
the settling of small dust grains toward the disk midplane.  This produces a
reduction in the scale height of the emitting particles and an overall decrease
in the level of excess emission across the disk SED.  Recently evidence for such
evolution in SEDs of young stars was obtained in Sptizer observations by
Sicilia-Aguilar et al. (2005).  These authors compared the SEDs of stars in
three stellar gourps of differing age, Taurus (1-2 Myr), Tr~37 (4 Myr) and
NGC~7160 (10 Myr).  In addition to the expected overall decrease in disk
fraction with age, they found a correlation of disk flux or infrared excess with
age, such that the stars that retained disks exhibited lower disk fluxes with
increasing age.  The disk fluxes at the shorter wavelengths, presumably from the
inner disk regions, seemed to decrease more rapidly than those at the longest
(i.e., $24\micron$) wavelength.  As mentioned earlier and shown in Figure
\ref{med_seds}, the median SED for IC~348 disk bearing stars is very similar to
the corresponding median SED for Taurus sources, suggesting similar disk
structure for corresponding sources in the two clouds.  Thus, even though IC~348
has a lower overall disk fraction, its surviving disks do not seem significantly
different in structure than those in Taurus.

\begin{figure}
\centering\includegraphics[angle=0,height=0.75\textheight]{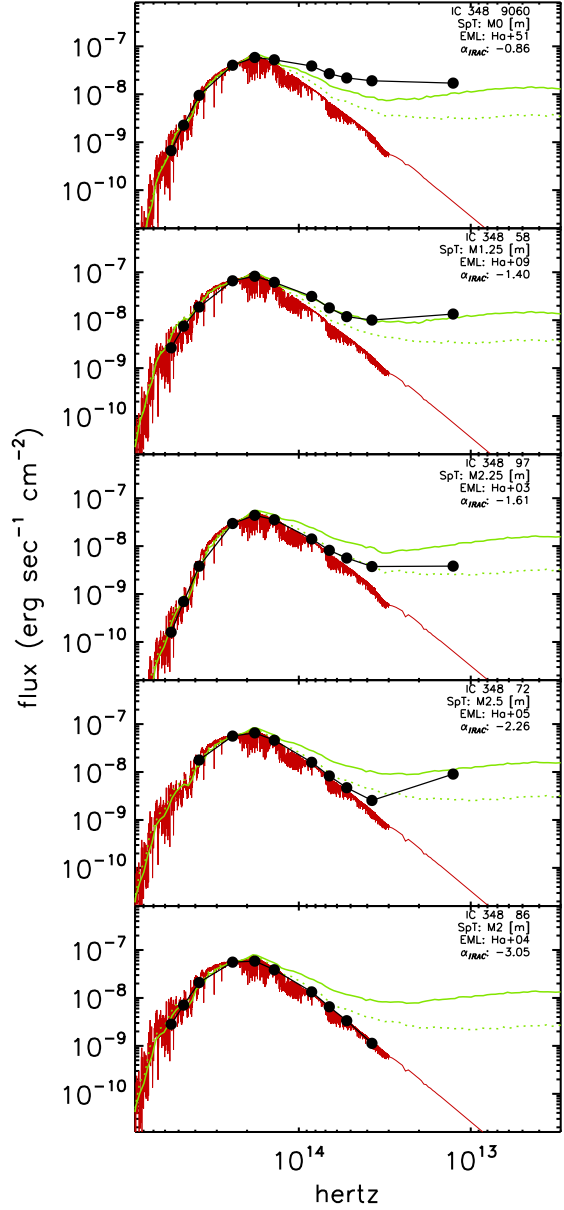}
\caption{Disk Evolution. SEDs for five M stars in the cluster exhibiting
differing levels of infrared excess and disk emission, ranging from a system
with a robust excess and highly flared disk structure through systems with
modest infrared excesses and flatter disk structures to a diskless star. Plotted
for comparison are the  corresponding  synthetic stellar photospheric models
(red jagged line) and two disk models whose scale heights differ. The  upper
disk model (solid line) has a scale height appropriate for a disk in hydrostatic
equilibrium, while the lower disk model (dotted line) has its scale height
reduced by a factor of 3. The lower disk scale height  corresponds to a
situation in which emitting dust has settled toward the  mid-plane. }
\label{sed_evolution}
\end{figure}

In Figure \ref{sed_evolution} we display the SEDs of four sources taken from our
atlas of SEDs as a representative sample of various SED shapes we find for cluster
stars.  These objects are of similar spectral type (M0 - M2) but exhibit different
levels of infrared excess due to circumstellar disks.  They range from a strongly
flared disk to a diskless photosphere.  These SEDs clearly represent systematically
varying disk structure for sources within the cluster.  For comparison appropriate
synthetic stellar photospheres and two disk models of differing scale height but
similar inclination (i.e., i $=$ 60$^o$) are also plotted for each star (see Appendix
1 for details).  Comparison with the model disks and photospheres suggests that the
decrease in disk emission from top to bottom is likely a result of a change in disk
scale height.  This, in turn, may be a sign of disk evolution due to dust settling.
If so, this suggests that the rate of disk evolution can vary within a single,
similarly aged, population of young stars.  It is interesting in this regard that we
find that for cluster members with spectral types earlier than M2, $70\%$ of the
disk-bearing stars have SEDs that closely match the prediction of the most flared
model disk (i.e., the upper disk model curve in Figure \ref{sed_evolution}) while
90\% of the disk-bearing stars with spectral types later than M4 exhibit SEDs that
lie below this model, coincident with or slightly above the prediction of the reduced
scale height disk model (i.e., the lower disk model curve).  Thus the disks around
the lowest mass, late-type stars are flatter and presumably more evolved than the
disks around the earlier type and solar mass stars.  This is interesting given that
the disk fraction around these later type stars is also lower or more evolved than
that for the solar mass stars in the cluster.  However, it is interesting to note
here that modeling of disks around brown dwarfs in Ophiuchus also suggested flatter
disk structure for such very low mass objects (Natta et al 2002).  The rapid disk
evolution times suggested by our observations are a challenge for theories of planet
formation which typically require periods of order 10$^7$ years or longer to build
Jupiter mass objects.  However, our data also suggest that the disk evolution time
scale is a function of stellar mass and may peak for stars similar in mass to the
sun.  Thus in IC~348 the conditions for planet building may be most favorable around
stars with masses similar to our sun.

We have so far considered the properties of optically thick disks which are
likely to be primordial, protoplanetary disks.  We have also identified a
population of anemic or optically thin disks in the cluster.  These disks appear
to be typically detected only at the longest wavelengths, $8$ and $24\micron\ $. 
The fraction of sources surrounded by such disks ranges from about 8\% for the
earliest type stars to 25\% for the later type stars.  However while such anemic
disks are a small portion of the disk-bearing early-type ($<$K6) stars, they
represent a large fraction (1/2) of the disk-bearing late-type ($>$M3) stars. 
This perhaps is another indication of more advanced evolution for the disks
around lower mass stars.  

Source 72, whose SED is shown in Figure \ref{sed_evolution}, is an interesting object
because it is an example of an anemic disk whose $24 \micron$\ (outer disk) emission is as
bright as that from the thick and flared primordial disks shown in the upper two panels of
Figure \ref{sed_evolution}.  This suggests the presence of a significant inner disk hole
with a fairly pronounced ring of outer disk emission.  Examination of the SEDs in the atlas
suggests that the evolution from a flared to a flat disk structure that is implied by the
arrangement of SEDs in Figure \ref{sed_evolution} is fairly typical in that the $24
\micron$\ flux almost always drops with the fluxes at shorter wavelengths.  This is
consistent with what one might expect if disks evolved more or less homologously (e.g.,
Wood et al.  2002).  Sources such as 72 represent an exception.  Their strong excess at $24
\micron$\ could either result from a sequence of inside-out evolution in which the dust in
the outer disk remained puffed up as the dust in the inner disk rapidly settled or perhaps
a sequence in which dust in the outer disk first settled to the mid-plane and then later
was puffed up again as the inner disk cleared.

Only 37\% of the anemic disk sources have been detected at $24\; \micron$.  These are the
sources identified in the IRAC-MIPS color-color diagram (Figure \ref{irac_mips}).  This low
detection fraction is the result of inadequate sensitivity at $24\; \micron$ and so these
particular sources are clearly the objects with the largest $24\; \micron$ excesses among
the anemic disk population.  Only 6 (or 33\%) of these sources exhibit $8-24$ \micron\
colors as extreme as Source 72.  The remaining anemic sources have $8-24$ \micron\ colors
which span a similar range as those of optically thick disks.  In addition, examination of
the individual SEDs shows that about half the anemic disks detected at 24 $\mu$m appear to
be thin at 24 $\mu$m.  These facts suggest that most disks may undergo substantial
evolution without developing significant inner holes, that is, the inner and outer regions
are more or less simultaneously being depleted by the evolutionary process.  However, it is
not clear from our observations whether the rates of inner and outer disk evolution could
still differ somewhat for these sources.

\begin{figure}
\plotone{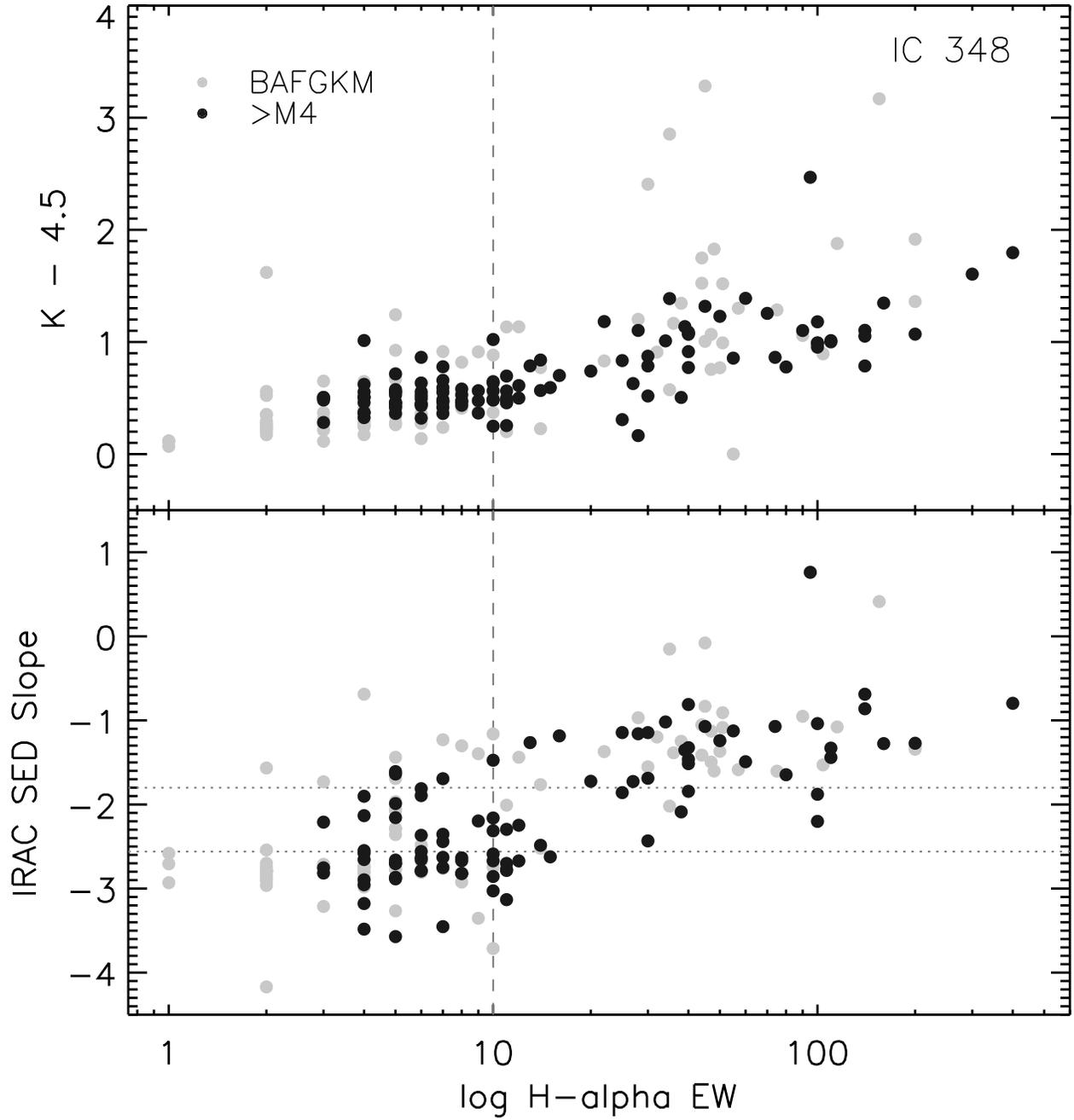}
\caption{Plot of the relation between H$\alpha$ equivalent width and \irac\ SED
(dereddened) power-law slope. Sources with disks are the strongest H$\alpha$
emitters indicating a link between accretion and the presence of a dusty,
optically thick circumstellar disk.}
\label{irac_vs_halpha}
\end{figure}

The data we have analyzed and discussed provides insights into inner (.1 - 5.0 AU) disk
evolution as traced by small dust grains.  It is not clear whether the evolution of the
gaseous component of the disk proceeds in a similar manner or on a similar timescale.  One
proxy for tracing the evolution of the gas is via measurements of disk accretion onto the
central star via observation of atomic emission lines.  In particular the H$\alpha$
emission is thought give an indication of the level of disk accretion onto the star.
Indeed, Hartigan et al.  (1990) found a correlation between optical veiling, which is an
indirect measure of accretion, and H$\alpha$ emission.  Moreover these authors also found a
correlation between veiling and infrared excess implying a correlation between the presence
of a dust disk and H$\alpha$ emission strength.  A similar result was found by Hartmann
(1998) who showed that stars with strong H$\alpha$ emission in Taurus tended to have large
K-L infrared excess.  In Figure \ref{irac_vs_halpha} we compare the H$\alpha$ equivalent
widths with the IRAC SED slopes of stars in the cluster.  Typically an equivalent width of
10 Angstroms is taken to be indicative of the presence of accretion and stars with
equivalent widths greater than this are designated as CTTS (classical T Tauri stars).
Sources with smaller equivalent widths are designated as WTTS (weak-lined T Tauri stars)
and are believed to be non-accreting objects or objects with very low accretion rates.
This boundary between accretion and chromospheric emission is somewhat arbitrary and likely
is not exactly appropriate for either very early or very late type stars.  (For example,
White and Basri (2003) suggest that for stars between K0 and K5 the boundary should be near
3 Angstroms; only three of stars in our sample earlier than K7 have equivalent widths
between 3-10 Angstroms.)  Figure \ref{irac_vs_halpha} shows a definite correlation between
H$\alpha$ equivalent width and SED slope.  In particular sources with slopes $(>-1.8)$
indicative of optically thick circumstellar disks have the strongest emission lines.  Of
the stars classified as CTTS and actively accreting objects, 68\% possess optically thick
disks and 11\% are associated with thin or anemic disks.  Of the stars classified as
non-accreting WTTS, 12\% are associated with robust optically thick disks and 22\% with
thin disks.  These data are not consistent with the notion that gaseous accretion disks
exist for significant periods of time after the small dust emitting grains have gone away.
This suggests that the gaseous and dust components of disks evolve on similar time scales.
Indeed, it appears more likely that dust disks can persist for at least some time after
gaseous accretion has been terminated.

\section{Summary and Conclusions}

We have combined \sst GTO and existing ground-based, optical-infrared observations
to obtain an accurate census and investigate the nature of the disk population of
the young embedded cluster, IC~348.  We have constructed optical-infrared SEDs for
all know members of the cluster spanning a spectral type range from A0 to M8. This
range includes stars at and somewhat below the hydrogen burning limit. From analysis
of these results we summarize our primary findings as follows:

\noindent
1)- From analysis of the optical-infrared SEDs of the cluster members we determine
the fraction of disk-bearing stars to be 50 $\pm$ 6\%. However only 30 $\pm$ 4\% 
of the member stars are surrounded by robust, optically thick or primordial disks
while 20\% of the member stars are surrounded by optically thin or anemic
disks. These measurements are lower than previous estimates for this cluster.
These measurements suggest that in the 2-3 Myr since the cluster formed, 70\% of
the stars have lost all or most of their primordial circumstellar disk material. 

\noindent 
2)- The disk fraction is a function of spectral type.  The lowest
disk frequency is found for early type stars.  Only 11 $\pm$ 8\% of stars earlier
than K6 possess optically thick disks.  The disk fraction peaks for stars of K6-M2
spectral types.  These PMS stars have masses similar to the sun and the fraction of
such stars with robust, thick disks is 47 $\pm$ 12\%.  For M2-M6 stars the overall
disk fraction is found to decline to a value of 28 $\pm$ 5\%.  Thus
it appears that disk longevity and the conditions for planet formation are most
favorable around the solar mass stars in this cluster.

\noindent 
3)- The optically thick disks around the lower mass, later type ($>$ M4)
stars in the cluster appear to be spatially flatter than those around the earlier
type stars.  This may indicate a greater degree of dust settling and more advanced
evolution in the disks associated with the late M population.

\noindent 
4)- A population of optically thin or anemic disks is identified.  The fraction
of stars accompanied by such disks ranges from 8 $\pm$ 8\% for early type stars to 26 $\pm$
5\% for M type stars.  However the fraction of disk-bearing stars with anemic disks is
small for stars earlier than K6, but is roughly 50\% for M stars.  These disks are almost
exclusively detected at the longest wavelengths $\lambda$ $\geq$ 8 $\micron$\ corresponding
to the outermost disk regions probed by the \sst observations.  We find that 37\% of the
anemic disk population is detected in the MIPS $24\micron$\ band.  In a few cases the $24
\micron$\ emission is as strong as that predicted for a flared, optically thick disk
suggesting that dust emission from the outer disk may persist or even strengthen in spite
of significant inner disk evolution and dissipation.  However, for most anemic disk sources
detected at 24 $\mu$m, the data are consistent with the possibility that the inner and
outer disk regions (as traced by \sst) evolve homologously and are more or less
simultaneously depleted of dusty material as they age.

\noindent 
5)- We find that the presence of an optically thick primordial disk is
correlated with gaseous accretion as measured by H$\alpha$ emission. A large
fraction (68\%) of stars classified as CTTS (by their measured H$\alpha$
equivalent widths) are surrounded by robust primordial disks.  Few such
stars are diskless. On  the other hand, 12\% of the stars classified as WTTS are
associated with thick disks, while 64\%  appear to be diskless.
This result suggests that active accretion and the evolution of
the dust disks  are strongly coupled. It is more likely for dust disks to
persist after active accretion ceases than for active accretion to persist
after the dust disks disappear.

\acknowledgments

This work is based on observations made with the {\it Spitzer Space Telescope}, which
is operated by the Jet Propulsion Laboratory, California Institute of Technology
under NASA contract 1407.  Support for this work was provided by NASA through
contract 1256790 issued by JPL/Caltech.  Support for the IRAC instrument was provided
by NASA through contract 960541 issued by JPL.  K.  L.  was supported by grant
NAG5-11627 from the NASA Long-Term Space Astrophysics program. We thank an anonymous
referee for a careful reading of the paper and thoughtful suggestions which improved 
its structure.

%


\appendix

\section{Atlas of Spectral Energy Distributions for IC~348}
\label{app:atlas}

We present an atlas of {\sl observed} spectral energy distributions for $307$ sources in
the IC~348 cluster, including 304 previously identified members.  The atlas is ordered by
member spectral type.  Seven members in very close binary pairs (9012, 9042, 9099, 22021,
30190, 30191, 30192) were excluded from this atlas since they only have photometry at
optical wavelengths.  Three sources classified as background stars based on their
subluminous location in optical-infrared color-magnitude diagrams have spectral energy
distributions suggesting they are in fact young stellar objects; we include these at the
end of the atlas.  Other than shifting the spectral energy distributions to a distance of
10pc (from 320pc assumed) the monochromatic fluxes shown in the atlas are unmodified from
the observations.  \mips\ $24\ \micron$ upper limits are shown as open circles; these upper
limits correpond to the net aperture flux within a $4.5\arcsec$ aperture centered on each
undetected star plus two times the noise, which includes photon and sky noise at that
location on the image.  This is the same definition for upper limits as used by the 2MASS
survey.  The references for the optical/near-infrared photometry used to create these SEDs
as well as the passband to flux density conversions are listed in Table
\ref{tab:passbands}.  In the optical, all the observed passbands were originally quoted as
being $VR_cI_c$.  All the near-infrared data were originally zeropoint calibrated to 2MASS,
hence we adopt the 2MASS flux calibration.

Each SED in the atlas is fit by a reddened model atmosphere.  The reddening law is a merger
of empirical laws from \citet{1989ApJ...345..245C} for $\lambda<1\micron, R_V = 3.1$) and
from a \sst\ study by \citet{2005ApJ...619..931I} for $1<\lambda<8\micron$.  The extinction
law (Case B, $R_V = 5.5$) of \citet{2001ApJ...548..296W} was used longward of $8\micron$.
A model atmosphere appropriate to each star was assigned based upon the IC~348 member's
spectral type and a conversion to effective temperature for pre-main sequence late-type
subgiants from \citet{1999ApJ...525..466L}.  Members without spectral types (e.g.,
continuum sources) were assigned a 3900K (K7) model atmosphere.  Model atmospheres were
taken from the Nextgen database for a surface gravity appropriate for PMS stars
$(\log\,g\,=4)$ except for the centrally located (and poorly fit) B5 source $\#1$ for which
we used a model atmosphere for 15600K dwarf taken from \citet{1993KurCD..13.....K}.  For
each star, the model atmosphere was reddened and fit to the observed SED using an iterative
filter convolution and $\chisq$ minimization routine; the fit range was from $0.5$ to
$2.0\;\micron$\ except for the latest spectral types, where we restricted the fit range to
$0.7$ to $2.0\;\micron$.  Normalization was to the $J$ passband.  The derived $\av$ have a
precision of 0.1 mag, though this is dependent upon the accuracy of the model atmospheres;
we list the derived $\av$ in Table \ref{tab:alphafits}.  Note that a more refined 5+
parameter model SED fit that includes effective temperature, reddening, surface gravity,
rotation and accretion luminosity might result in improved fits to the IC~348 members' blue
optical SEDs.  Including the effects of rapid rotation and accretion related blue excess
(veiling) in
the optical, might clarify whether or not some of the poor matches in the optical SED were
due to these effects or a lack of precision in the assigned spectral type. However,
such fits are beyond the scope of the present investigation.

For stars later than F type the SEDs are compared to two disk models both
assuming a ``typical'' inclination, i.e., $\cos\,i\:=0.5$.  The stellar
parameters of the disk models were varied as a function of each star's spectral
type. Specifically, the disk models we show appropriately correspond to G3, K2,
K7, M1, M2 or M5 photospheres; the latter star+disk models were used for all
members later than M5, including the brown dwarf SEDs.  The disk models shown
are not fit to the observed SEDs; they are simply reddened by the
\chisq~minimized $\av$ derived above and normalized to $J$ band using filter
convolution.  Brief astrographical data is given for each star in the atlas,
including its catalog number, spectral type, membership criteria and emission
line properties.  The reddening $(\av)$ to a star as measured by the fit model
atmosphere is listed as is the observed slope of the \irac\ spectral energy
distribution $(\alpha_{IRAC})$. These values are also listed in Table
\ref{tab:alphafits}.  

\noindent We note a few individual sources:

\paragraph{746, 2096, 212}  These sources were previously thought to be
background to the IC~348 cluster.  Though $212$ clearly appears to have an
evolved ``anemic'' type disk and could be an interloper from a previous epoch of
star formation (Perseus OB2, for example) the other two M5-M6 sources have very
strong disk signatures.  That their bolometric luminosities are more than 1
order  of magnitude lower than comparable M5-M6 stars suggests that they are
probably seen nearly edge-on \citep{1999ApJ...527..893D}, explaining why they
were not included in previous census.

\paragraph{276, 203, 435}  The IRAC-MIPS SEDs of these sources suggest that they
are the best candidate IC~348 Class I sources with known spectral types. That they
also have very low measured $\av$ suggests their optical colors are contaminated
by scattered light (cf. 203). 

\section{Median Observed Spectral Energy Distributions}
\label{app:median}

We know a priori that the observed spectral energy distributions of very young 
stellar objects probably do not correspond to those of typical main sequence
dwarfs for three straightforward reasons: 1) very young PMS star SEDs are
distorted by the reradiation of stellar light by a circumstellar disk, 2) their
SEDs can be distorted by rapid rotation and/or active chromospheres
\citep{2003AJ....126..833S} and 3) the large radii instrinsic to contracting PMS
stars requires that they have more giant-like surface gravities.  Empirically we
know that the latest spectral classes correspond to subgiant gravities
\citep{1999ApJ...525..466L} and display unusual optical colors
\cite{2000AJ....119.3026R}.  Sorting out the shape of the spectral energy
distribution of young stars requires a set of sources with refined photometry
covering the optical and mid-IR and with excellent spectral types.  The ensemble
of data we have collected for IC~348 satisfies those  requirements probably
better than any other young stellar cluster or star forming region; further, the
\sst data presented here allows us to parse diskless stellar photospheres from 
disked stars with excellent precision.  Typical photospheric SEDs for IC~348 stars
would provide a good starting place for detailed tests of model atmospheres.

Another valuable reason for deriving the typical SEDs of IC~348 stars is to
model the evolution of circumstellar disks. Recent comparisons of star-disk
spectral energy distribution as a function of age have sketched a steady
decrease in the disk flux in the IRAC bands for accreting classical type T-Tauri
stars \citep{saetal05} over a 10 Myr period. Placing the typical SEDs from
IC~348 into such a sequence may prove useful for refining both time and
environmental variations in inner disk properties; moreover, no other young
cluster dataset probes a statistically significant population of sources down to
and below the hydrogen burning limit as we have collected for IC~348.  

\begin{figure} 
\plottwo{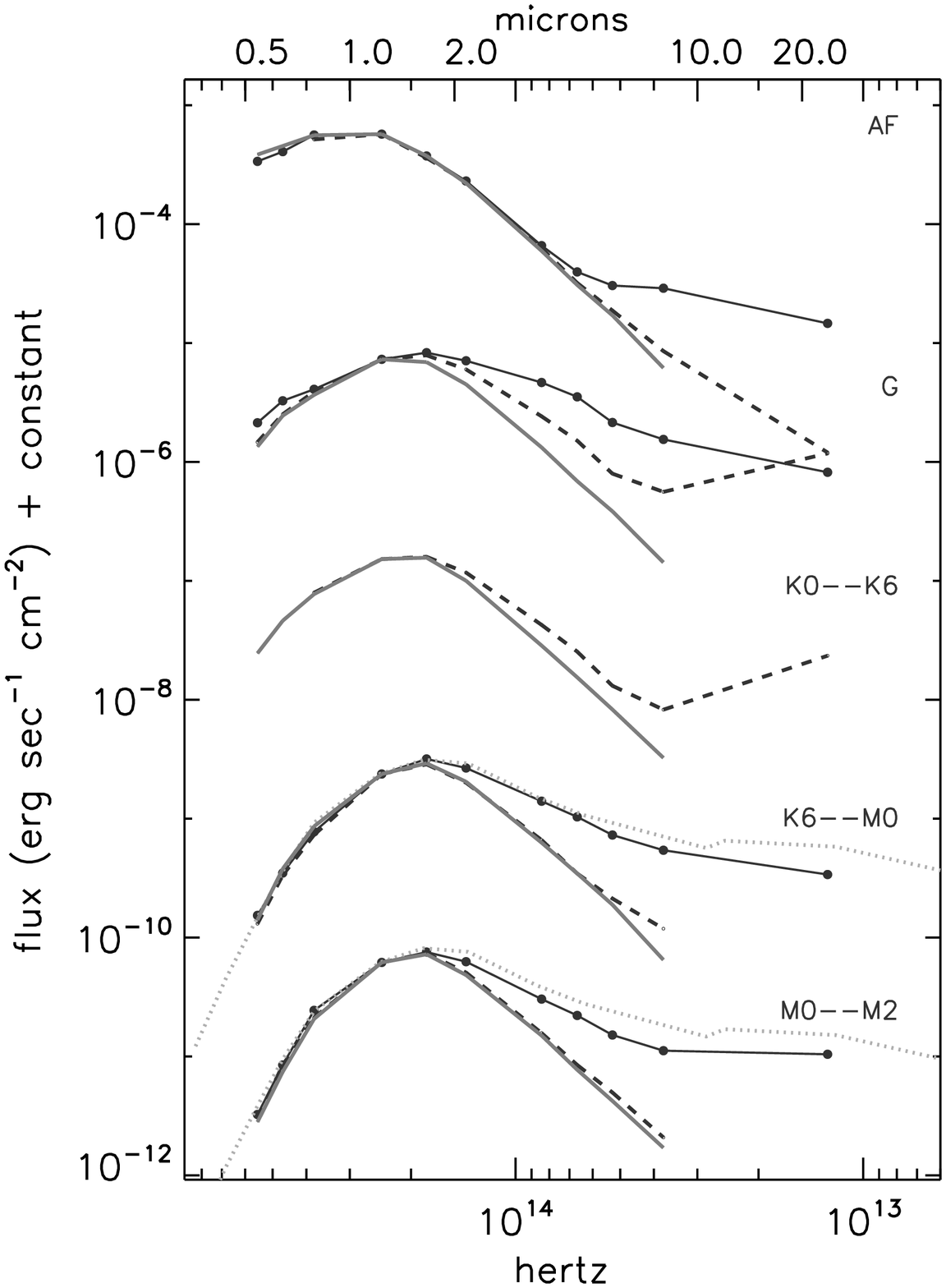}{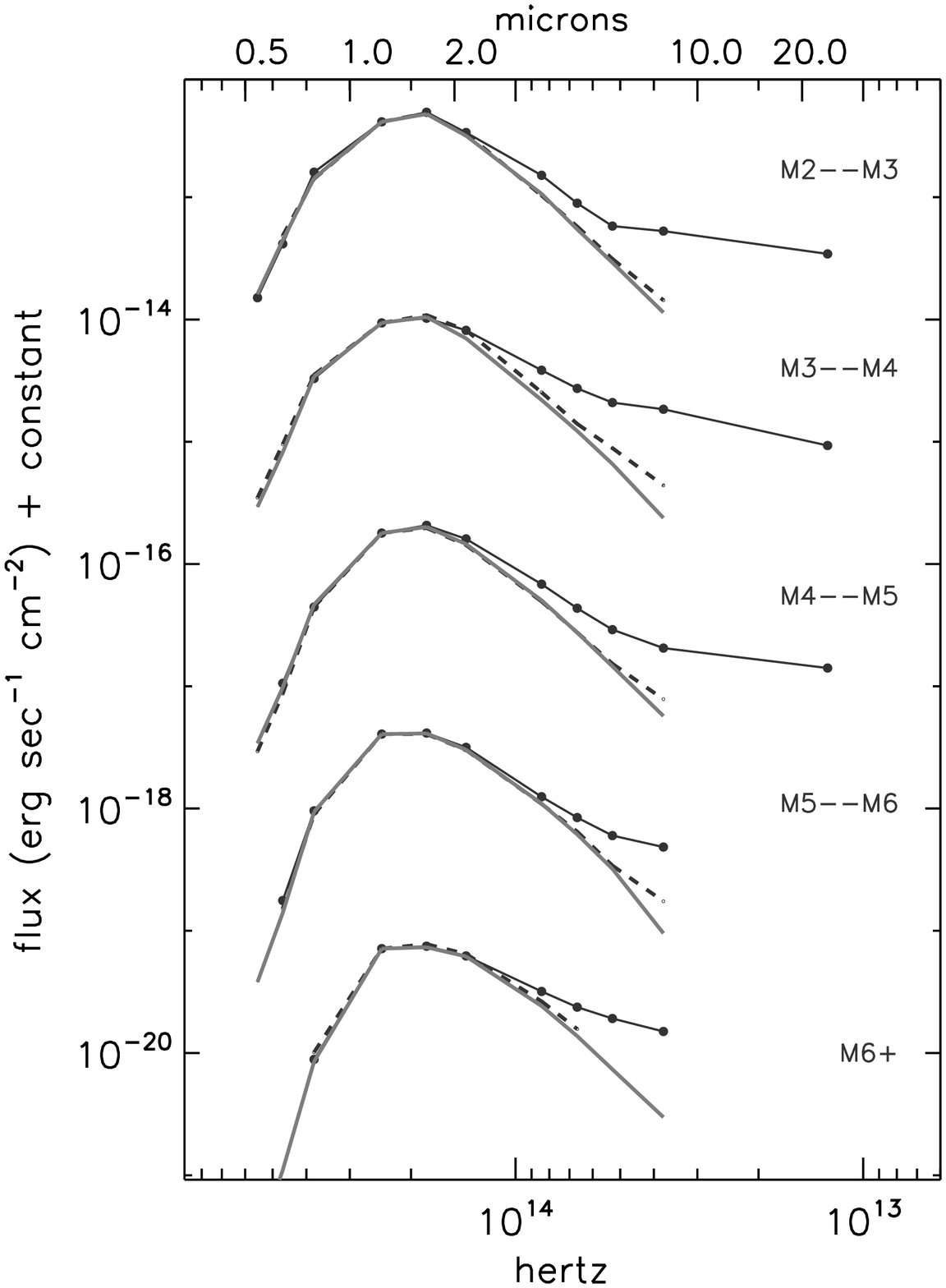}
\caption{Median IC~348 SEDs for optically thick flared disks (solid colored
line),  anemic disks (dashed colored line) and diskless stellar photospheres
(grey solid line) for the various spectral type bins.  The median SED for K0-M2
classical T-Tauri stars in the  Taurus cloud \citep{1999ApJ...527..893D} is also
compared to the  appropriate IC~348 SEDs. Note that these are the median
\emph{typical, observed} SEDs as a function of spectral type in IC~348 and thus
do not correspond to $\av\neq0$; see Appendix \ref{app:median}.}

\label{med_seds} 
\end{figure}

Given our goal of providing median SEDs that correspond to the
typical observed star or star/disk system in the 2-3 Myr IC~348 cluster, we
present in Figure \ref{med_seds} median spectral energy distributions for
diskless stars in IC~348, members with anemic disks and members with thick
disks, while breaking the ensemble population into ten spectral type bins. 
Additionally and for the reasons listed above, deriving and publishing de-reddened
$(\av=0)$ median SEDs for IC~348 would be an imprecise goal since the only obvious 
templates for dereddening would be dwarf SEDs. Instead we used the observed
fact that the reddenings toward stars in IC~348 are typically low $(\av\sim2)$
to derive median \emph{typical, observed} SEDs for each spectral type range.

Our prescription for deriving these median SEDs is as follows. Simply, a median
filter was passed over the set of observed SEDs in a spectral type range to
essentially select that ``star'' whose SED is typical of that spectral class. To
parse the different star/disk classes, each observed SED in the spectral subclass
was reddened {\bf or} de-reddened to match the optical portion of this typical
SED; then they were normalized to match the typical SED's $J$ band flux. Next,
the IRAC SED slopes were measured and the three classifications of star/disk
systems (star, anemic and thick) were parsed according to the
$\alpha_{\irac}$~ranges given in Section 3.1. Note that these are not the origin
of the  observed and de-reddened  $\alpha_{\irac}$ listed in Table
\ref{tab:alphafits}; see Appendix \ref{app:atlas}. Finally, each set of
star/disk normalized SEDs were median filtered by bandpass to yield the median
SEDs plotted in Figure \ref{med_seds} and the values given in Table
\ref{tab:median}.  In this Table we list the relevant numerical data for these
median SEDs (for each spectral type bin and star/disk system), including the
median magnitudes and fluxes for each SED, the $1\sigma$ dispersion in the
fluxes and the number of data points that went into each median SED point. 

Finally, we believe that the resulting median observed SEDs are
reddened by approximately $\av \sim 2.5$; for example, the Taurus median SED
\citep{1999ApJ...527..893D}  compared to the IC~348 K6--M0 and M0--M2 spectral
class SEDs in Figure \ref{med_seds} were arbitrarily reddened by $\av=2.5$;
yielding the excellent SED agreement shown in those panels.



\clearpage

\clearpage

\clearpage

\clearpage

\clearpage

\clearpage

\clearpage

\clearpage


\begin{figure}
\figurenum{A-1}
   \centering
     \begin{minipage}[c]{0.75\textwidth}
     \centering \includegraphics[angle=0,width=\textwidth]{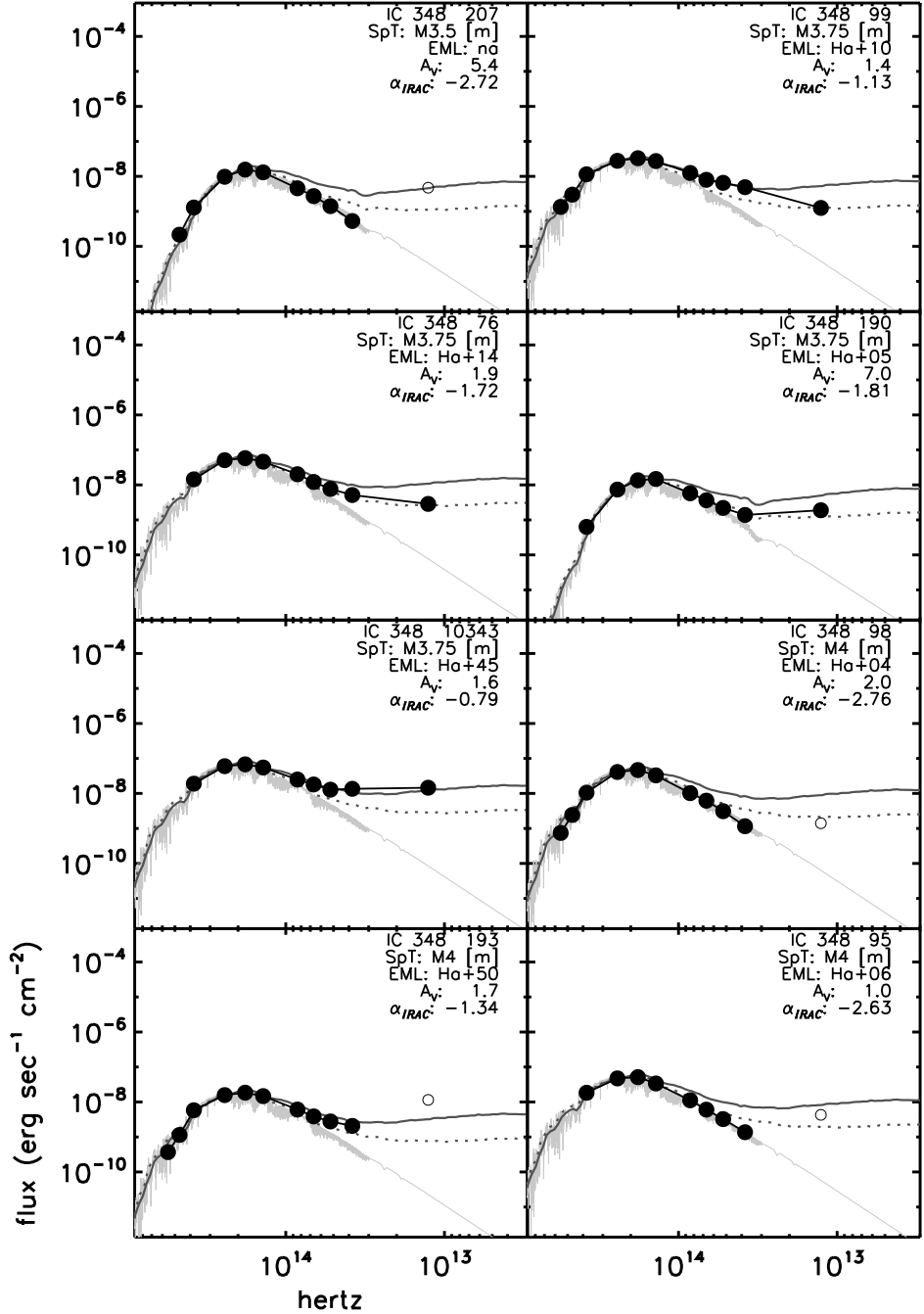}
     \end{minipage}%
\caption{Atlas of Observed Spectral Energy Distributions  for members of the 
2-3 Myr IC~348 cluster. SAMPLE PAGE. 
Astrographical ($\av$, $\alpha_{IRAC}$, Spectral type, etc)
information is given for each member. The best fit appropriately reddened model
atmosphere is overplotted as are two reddened disk models with different disk scale
heights. The underlying photosphere of the disk models were varied according
to the spectral type of the IC~348 member. Filled circles represent the detected
fluxes and open circles upper limits to the observed fluxes. Error bars are also
plotted but in almost all cases are smaller than the plotted circles.
See Appendix \ref{app:atlas} and text. (Complete catalog of SEDs and complete electronic
tabular material can be obtained at: 
http://www.cfa.harvard.edu/$\sim$clada/pubs\_html/ic348\_spitzer.html).
}
\label{atlas}
\end{figure}



\clearpage

\begin{deluxetable}{rllrrrrrrrrrrcc}
\tablecolumns{15}
\tablewidth{0pt}

\tabletypesize{\tiny}
\tablecaption{\sst\ IRAC/MIPS IC~348 Catalog
\label{tab:catalog}}

\tablehead{
\colhead{No.}  &
\colhead{R.A.}  &
\colhead{Dec.}  &
\multicolumn{8}{c}{\irac\ Photometry} &
\multicolumn{2}{c}{\mips\ Photometry} &
\multicolumn{2}{c}{Blend Flags} \\

\colhead{}        &
\colhead{(J2000)} &
\colhead{(J2000)} &
\colhead{\SIa}    &
\colhead{err}     &
\colhead{\SIb}    &
\colhead{err}     &
\colhead{\SIc}    &
\colhead{err}     &
\colhead{\SId}    &
\colhead{err}     &
\colhead{\SMa}    &
\colhead{err\ \tablenotemark{(a)}}    &
\colhead{\irac\ \tablenotemark{(b)}}  &
\colhead{\mips\ \tablenotemark{(c)}}   
}
\startdata
      1 &  03:44:34.212 &   32:09:46.69 &     6.74 &     0.01 &     6.54 &     0.02 &     6.58 &     0.02 &     6.50 &     0.03 &     0.89 &    -9.00 &         &         \\
      2 &  03:44:35.364 &   32:10:04.58 &     7.09 &     0.02 &     6.81 &     0.02 &     6.46 &     0.04 &     5.82 &     0.04 &     3.20 &     0.03 &         &         \\
      3 &  03:44:50.649 &   32:19:06.75 &     7.53 &     0.01 &     7.61 &     0.03 &     7.47 &     0.04 &     7.50 &     0.04 &     7.08 &     0.05 &         &         \\
      4 &  03:44:31.188 &   32:06:22.09 &     7.79 &     0.01 &     7.74 &     0.02 &     7.66 &     0.02 &     7.73 &     0.03 &     4.38 &    -9.00 &         &         \\
      5 &  03:44:26.027 &   32:04:30.41 &     6.97 &     0.01 &     6.52 &     0.01 &     6.32 &     0.02 &     5.63 &     0.02 &     2.76 &     0.03 &         &         \\
      6 &  03:44:36.941 &   32:06:45.37 &     7.92 &     0.02 &     7.70 &     0.01 &     7.53 &     0.05 &     7.12 &     0.03 &     5.37 &     0.04 &         &         \\
      7 &  03:44:08.476 &   32:07:16.50 &     8.54 &     0.01 &     8.60 &     0.01 &     8.53 &     0.03 &     8.48 &     0.03 &     6.95 &     0.05 &         &         \\
      8 &  03:44:09.152 &   32:07:09.33 &     8.61 &     0.01 &     8.62 &     0.01 &     8.52 &     0.03 &     8.33 &     0.04 &     6.96 &     0.06 &         &         \\
      9 &  03:44:39.178 &   32:09:18.35 &     8.47 &     0.01 &     8.41 &     0.02 &     8.35 &     0.03 &     8.50 &     0.10 &     4.14 &    -9.00 &         &         \\
     10 &  03:44:24.664 &   32:10:15.04 &     8.71 &     0.02 &     8.67 &     0.01 &     8.64 &     0.03 &     8.68 &     0.04 &     4.81 &    -9.00 &         &         \\
     11 &  03:45:07.965 &   32:04:02.09 &     8.59 &     0.01 &     8.63 &     0.01 &     8.44 &     0.06 &     8.44 &     0.01 &     8.32 &     0.04 &         &         \\
     13 &  03:43:59.641 &   32:01:54.17 &     7.22 &     0.01 &     6.42 &     0.01 &     5.83 &     0.03 &     4.98 &     0.04 &     2.11 &     0.03 &         &         \\
     15 &  03:44:44.716 &   32:04:02.72 &     8.52 &     0.01 &     8.17 &     0.03 &     7.81 &     0.03 &     7.17 &     0.02 &     4.11 &     0.03 &         &         \\
     16 &  03:44:32.743 &   32:08:37.46 &     9.36 &     0.01 &     9.37 &     0.03 &     9.22 &     0.06 &     9.27 &     0.12 &     5.42 &    -9.00 &         &         \\
     17 &  03:44:47.724 &   32:19:11.95 &     9.25 &     0.01 &     9.11 &     0.01 &     9.21 &     0.05 &     9.12 &     0.08 &     6.13 &    -9.00 &         &         \\
\enddata

\tablenotetext{(a)}{If the \mips\  $24\micron$ magnitude error
corresponds to $-9$ then the \mips\  $24\micron$ flux is a $95\%$
upper limit. See text.}
\tablenotetext{(b)}{\irac\ blend flag.}
\tablenotetext{(c)}{\mips\ blend flag.}

\tablecomments{The complete version of this table is in the electronic
edition of the Journal.  The printed edition contains only this sample.}

\tablecomments{This table is available only on-line as a machine-readable table.}

\end{deluxetable}  


\clearpage

\begin{deluxetable}{rlrrrrrcr}
\tablecolumns{9}
\tablewidth{0pt}

\tabletypesize{\footnotesize}
\tablecaption{SED derived \irac\ $\alpha$ and $\av$
\label{tab:alphafits}}

\tablehead{
\colhead{No.}  &
\colhead{SpT} &
\colhead{$\av$\tablenotemark{(a)}} &
\multicolumn{5}{c}{$\alpha_{\irac}$~Fits\tablenotemark{(b)}} &
\colhead{Disk} \\

\colhead{}        &
\colhead{}        &
\colhead{[mag}    &
\colhead{$\alpha_{0}$}   &
\colhead{$1\sigma$}      &
\colhead{$\alpha_{d}$}   &
\colhead{$1\sigma$}      &
\colhead{N(bands)}       &
\colhead{Type\tablenotemark{(c)}}       

}
\startdata
      1 &      B5 &    3.1 &   -2.638 &    0.102 &   -2.708 &    0.091 &    4 &     STAR \\
      2 &      A2 &    3.2 &   -1.396 &    0.127 &   -1.468 &    0.143 &    4 &    THICK \\
      3 &      A0 &    3.9 &   -2.794 &    0.110 &   -2.883 &    0.115 &    4 &     STAR \\
      4 &      F0 &    2.3 &   -2.786 &    0.091 &   -2.838 &    0.082 &    4 &     STAR \\
      5 &      G8 &    7.7 &   -1.389 &    0.160 &   -1.563 &    0.177 &    4 &    THICK \\
      6 &      G3 &    3.5 &   -1.972 &    0.079 &   -2.051 &    0.094 &    4 &   ANEMIC \\
      7 &      A0 &    1.7 &   -2.788 &    0.071 &   -2.826 &    0.078 &    4 &     STAR \\
      8 &      A2 &    1.6 &   -2.532 &    0.101 &   -2.569 &    0.109 &    4 &   ANEMIC \\
      9 &      G8 &    5.3 &   -2.894 &    0.134 &   -3.014 &    0.112 &    4 &     STAR \\
     10 &      F2 &    2.1 &   -2.827 &    0.054 &   -2.875 &    0.044 &    4 &     STAR \\
     11 &      G4 &    6.0 &   -2.653 &    0.114 &   -2.789 &    0.121 &    4 &     STAR \\
     13 &    M0.5 &   19.0 &   -0.315 &    0.115 &   -0.744 &    0.056 &    4 &    THICK \\
     15 &    M0.5 &    2.5 &   -1.325 &    0.088 &   -1.382 &    0.100 &    4 &    THICK \\
     16 &      G6 &    3.0 &   -2.724 &    0.098 &   -2.792 &    0.095 &    4 &     STAR \\
     17 &      A4 &    5.0 &   -2.772 &    0.098 &   -2.885 &    0.091 &    4 &     STAR \\
\enddata

\tablenotetext{(a)}{$\av$ derived from SED fitting with model atmospheres; see
text.}
\tablenotetext{(b)}{\irac\ $\alpha$ fits. $\alpha_{0}$: Observed slope;
$\alpha_{d}$: Dereddened slope }
\tablenotetext{(c)}{Disk Type based upon $\alpha_{d}$ }

\tablecomments{The complete version of this table is in the electronic
edition of the Journal.  The printed edition contains only this sample.}

\tablecomments{This table is available only on-line as a machine-readable table.}

\end{deluxetable}


\clearpage

\begin{deluxetable}{lrrrrrrrrrrr}
\tablecolumns{12}
\tablewidth{625pt}
\rotate

\tabletypesize{\scriptsize}
\tablecaption{Median IC~348 Star/Disk Spectral Energy Distributions
\label{tab:median}}

\tablehead{
\colhead{Measurement}  &
\colhead{\Vl}  &
\colhead{\Rc}  &
\colhead{\Ic}  &
\colhead{\Jtm}  &
\colhead{\Htm}  &
\colhead{\Ks}  &
\colhead{\SIa}  &
\colhead{\SIb}  &
\colhead{\SIc}  &
\colhead{\SId}  &
\colhead{\SMa}  
}
\startdata
\cutinhead{AF Spectral Types}
\sidehead{\bf Stars}
             magnitudes &          11.588 &         \nodata &          10.304 &           9.343 &           8.997 &           8.821 &           8.749 &           8.737 &           8.655 &           8.727 &           7.194 \\
      median [log] flux &          -6.312 &         \nodata &          -6.148 &          -6.139 &          -6.322 &          -6.552 &          -7.121 &          -7.408 &          -7.663 &          -8.106 &          -8.902 \\
$1\sigma;\;(N_{stars})$ &       -7.027(3) &      \nodata(0) &       -7.366(6) &      \nodata(7) &       -8.009(7) &       -8.047(7) &       -8.416(7) &       -8.665(7) &       -8.850(7) &       -9.179(7) &       -9.393(4) \\
\sidehead{\bf Anemic Disks}
             magnitudes &         \nodata &         \nodata &          10.392 &           9.343 &           9.040 &           8.797 &           8.696 &           8.695 &           8.554 &           8.373 &           6.982 \\
      median [log] flux &         \nodata &         \nodata &          -6.184 &          -6.139 &          -6.339 &          -6.542 &          -7.100 &          -7.391 &          -7.623 &          -7.964 &          -8.817 \\
$1\sigma;\;(N_{stars})$ &      \nodata(0) &      \nodata(0) &      \nodata(1) &      \nodata(1) &      \nodata(1) &      \nodata(1) &      \nodata(1) &      \nodata(1) &      \nodata(1) &      \nodata(1) &      \nodata(1) \\
\sidehead{\bf Thick Disks}
             magnitudes &          11.731 &          11.133 &          10.299 &           9.343 &           9.000 &           8.776 &           8.640 &           8.463 &           8.028 &           7.051 &           4.267 \\
      median [log] flux &          -6.368 &          -6.287 &          -6.147 &          -6.139 &          -6.323 &          -6.533 &          -7.077 &          -7.298 &          -7.413 &          -7.435 &          -7.731 \\
$1\sigma;\;(N_{stars})$ &      \nodata(1) &      \nodata(1) &       -7.195(2) &      \nodata(2) &       -8.011(2) &       -7.800(2) &       -8.410(2) &       -8.107(2) &       -8.629(2) &       -8.091(2) &       -8.201(2) \\

\cutinhead{G Spectral Types}
\sidehead{\bf Stars}
             magnitudes &          13.843 &          12.801 &          11.881 &          10.195 &           9.451 &           9.168 &           8.998 &           8.979 &           8.892 &           8.932 &           8.681 \\
      median [log] flux &          -7.213 &          -6.954 &          -6.779 &          -6.480 &          -6.504 &          -6.690 &          -7.220 &          -7.504 &          -7.758 &          -8.188 &          -9.496 \\
$1\sigma;\;(N_{stars})$ &       -8.484(3) &       -8.127(3) &       -7.970(6) &      \nodata(6) &       -8.007(6) &       -7.988(6) &       -8.410(6) &       -8.654(6) &       -8.955(6) &       -9.092(6) &       -9.629(3) \\
\sidehead{\bf Anemic Disks}
             magnitudes &          13.749 &          12.771 &          11.813 &          10.195 &           9.308 &           8.853 &           8.345 &           8.128 &           8.100 &           7.449 &           3.118 \\
      median [log] flux &          -7.176 &          -6.942 &          -6.752 &          -6.480 &          -6.446 &          -6.564 &          -6.959 &          -7.164 &          -7.441 &          -7.595 &          -7.271 \\
$1\sigma;\;(N_{stars})$ &      \nodata(1) &       -8.069(2) &       -7.685(2) &      \nodata(2) &       -7.265(2) &       -7.124(2) &       -7.339(2) &       -7.544(2) &       -7.911(2) &       -7.922(2) &       -7.293(2) \\
\sidehead{\bf Thick Disks}
             magnitudes &          13.343 &          12.488 &          11.763 &          10.195 &           9.252 &           8.672 &           7.634 &           7.207 &           7.024 &           6.347 &           3.515 \\
      median [log] flux &          -7.013 &          -6.829 &          -6.732 &          -6.480 &          -6.424 &          -6.492 &          -6.675 &          -6.795 &          -7.011 &          -7.154 &          -7.430 \\
$1\sigma;\;(N_{stars})$ &      \nodata(1) &      \nodata(1) &      \nodata(1) &      \nodata(1) &      \nodata(1) &      \nodata(1) &      \nodata(1) &      \nodata(1) &      \nodata(1) &      \nodata(1) &      \nodata(1) \\

\cutinhead{K0--K6 Spectral Types}
\sidehead{\bf Stars}
             magnitudes &          15.154 &          14.078 &          13.035 &          11.360 &          10.528 &          10.263 &          10.131 &          10.069 &          10.026 &          10.007 &         \nodata \\
      median [log] flux &          -7.738 &          -7.465 &          -7.241 &          -6.946 &          -6.934 &          -7.128 &          -7.673 &          -7.941 &          -8.212 &          -8.618 &         \nodata \\
$1\sigma;\;(N_{stars})$ &       -8.786(9) &       -8.412(9) &      -8.256(13) &     \nodata(13) &      -8.682(12) &      -8.682(13) &      -9.009(13) &      -9.192(13) &      -9.347(13) &      -9.688(13) &      \nodata(0) \\
\sidehead{\bf Anemic Disks}
             magnitudes &         \nodata &         \nodata &          13.004 &          11.360 &          10.509 &          10.091 &           9.695 &           9.528 &           9.525 &           8.994 &           4.337 \\
      median [log] flux &         \nodata &         \nodata &          -7.228 &          -6.946 &          -6.927 &          -7.060 &          -7.499 &          -7.724 &          -8.012 &          -8.213 &          -7.759 \\
$1\sigma;\;(N_{stars})$ &      \nodata(0) &      \nodata(0) &      \nodata(1) &      \nodata(1) &      \nodata(1) &      \nodata(1) &      \nodata(1) &      \nodata(1) &      \nodata(1) &      \nodata(1) &      \nodata(1) \\
\sidehead{\bf Thick Disks}
             magnitudes &         \nodata &         \nodata &         \nodata &         \nodata &         \nodata &         \nodata &         \nodata &         \nodata &         \nodata &         \nodata &         \nodata \\
      median [log] flux &         \nodata &         \nodata &         \nodata &         \nodata &         \nodata &         \nodata &         \nodata &         \nodata &         \nodata &         \nodata &         \nodata \\
$1\sigma;\;(N_{stars})$ &      \nodata(0) &      \nodata(0) &      \nodata(0) &      \nodata(0) &      \nodata(0) &      \nodata(0) &      \nodata(0) &      \nodata(0) &      \nodata(0) &      \nodata(0) &      \nodata(0) \\

\cutinhead{K6--M0 Spectral Types}
\sidehead{\bf Stars}
             magnitudes &          16.812 &          15.346 &          13.969 &          11.931 &          10.903 &          10.541 &          10.327 &          10.247 &          10.181 &          10.306 &          10.031 \\
      median [log] flux &          -8.401 &          -7.972 &          -7.614 &          -7.174 &          -7.085 &          -7.239 &          -7.752 &          -8.012 &          -8.274 &          -8.737 &         -10.036 \\
$1\sigma;\;(N_{stars})$ &       -9.315(3) &       -8.884(4) &       -8.489(5) &      \nodata(5) &       -8.324(5) &       -8.404(5) &       -8.736(5) &       -8.956(5) &       -9.092(5) &       -9.427(5) &      \nodata(1) \\
\sidehead{\bf Anemic Disks}
             magnitudes &          16.894 &          15.479 &          14.148 &          11.931 &          10.924 &          10.561 &          10.271 &          10.246 &          10.051 &           9.651 &         \nodata \\
      median [log] flux &          -8.434 &          -8.025 &          -7.686 &          -7.174 &          -7.093 &          -7.248 &          -7.730 &          -8.011 &          -8.222 &          -8.475 &         \nodata \\
$1\sigma;\;(N_{stars})$ &      \nodata(1) &      \nodata(1) &      \nodata(1) &      \nodata(1) &      \nodata(1) &      \nodata(1) &      \nodata(1) &      \nodata(1) &      \nodata(1) &      \nodata(1) &      \nodata(0) \\
\sidehead{\bf Thick Disks}
             magnitudes &          16.719 &          15.435 &          14.084 &          11.931 &          10.812 &          10.252 &           9.455 &           9.055 &           8.715 &           8.001 &           4.990 \\
      median [log] flux &          -8.364 &          -8.007 &          -7.660 &          -7.174 &          -7.048 &          -7.124 &          -7.403 &          -7.535 &          -7.688 &          -7.816 &          -8.020 \\
$1\sigma;\;(N_{stars})$ &       -9.516(5) &       -8.853(5) &       -8.352(7) &      \nodata(7) &       -8.312(7) &       -7.986(7) &       -7.916(7) &       -7.924(7) &       -8.182(7) &       -8.492(7) &       -8.345(7) \\

\cutinhead{M0--M2 Spectral Types}
\sidehead{\bf Stars}
             magnitudes &          17.725 &          16.277 &          14.682 &          12.557 &          11.583 &          11.269 &          11.046 &          11.036 &          10.970 &          10.921 &         \nodata \\
      median [log] flux &          -8.766 &          -8.344 &          -7.900 &          -7.425 &          -7.357 &          -7.531 &          -8.040 &          -8.327 &          -8.590 &          -8.984 &         \nodata \\
$1\sigma;\;(N_{stars})$ &      -10.277(5) &       -9.636(5) &      -8.775(12) &     \nodata(13) &      -8.882(12) &      -8.926(12) &      -9.263(12) &      -9.532(12) &      -9.766(12) &     -10.139(12) &      \nodata(0) \\
\sidehead{\bf Anemic Disks}
             magnitudes &          17.633 &          16.138 &          14.657 &          12.557 &          11.561 &          11.216 &          10.993 &          10.936 &          10.793 &          10.704 &           4.974 \\
      median [log] flux &          -8.729 &          -8.289 &          -7.890 &          -7.425 &          -7.348 &          -7.510 &          -8.019 &          -8.287 &          -8.519 &          -8.896 &          -8.013 \\
$1\sigma;\;(N_{stars})$ &      -10.670(3) &       -9.384(3) &       -8.817(4) &      \nodata(4) &       -9.027(4) &       -9.012(4) &       -8.963(4) &       -9.098(4) &       -9.352(4) &       -9.285(4) &      \nodata(1) \\
\sidehead{\bf Thick Disks}
             magnitudes &          17.568 &          16.135 &          14.503 &          12.557 &          11.537 &          10.989 &          10.280 &           9.895 &           9.584 &           8.880 &           5.433 \\
      median [log] flux &          -8.704 &          -8.287 &          -7.828 &          -7.425 &          -7.338 &          -7.419 &          -7.733 &          -7.871 &          -8.035 &          -8.167 &          -8.197 \\
$1\sigma;\;(N_{stars})$ &       -9.436(7) &       -8.933(7) &      -8.351(15) &     \nodata(15) &      -8.503(14) &      -8.043(14) &      -7.956(14) &      -8.072(13) &      -8.586(11) &      -8.444(12) &      -8.428(12) \\

\cutinhead{M2--M3 Spectral Types}
\sidehead{\bf Stars}
             magnitudes &          18.151 &          16.658 &          14.906 &          12.816 &          11.854 &          11.549 &          11.241 &          11.236 &          11.198 &          11.184 &         \nodata \\
      median [log] flux &          -8.937 &          -8.497 &          -7.989 &          -7.528 &          -7.465 &          -7.643 &          -8.118 &          -8.407 &          -8.681 &          -9.088 &         \nodata \\
$1\sigma;\;(N_{stars})$ &      -10.015(8) &       -9.552(8) &      -8.864(11) &     \nodata(12) &      -8.965(12) &      -9.098(12) &      -9.371(12) &      -9.639(12) &      -9.696(12) &     -10.203(12) &      \nodata(0) \\
\sidehead{\bf Anemic Disks}
             magnitudes &          18.014 &          16.565 &          14.919 &          12.816 &          11.836 &          11.500 &          11.274 &          11.180 &          11.120 &          10.931 &           5.773 \\
      median [log] flux &          -8.882 &          -8.459 &          -7.994 &          -7.528 &          -7.458 &          -7.623 &          -8.131 &          -8.385 &          -8.650 &          -8.987 &          -8.333 \\
$1\sigma;\;(N_{stars})$ &       -9.908(2) &       -9.772(4) &       -9.077(7) &      \nodata(7) &       -8.888(7) &       -8.739(7) &       -8.692(7) &       -8.820(7) &       -8.978(7) &       -9.429(7) &      \nodata(1) \\
\sidehead{\bf Thick Disks}
             magnitudes &          18.228 &          16.733 &          14.783 &          12.816 &          11.817 &          11.482 &          10.862 &          10.704 &          10.450 &           9.518 &           6.463 \\
      median [log] flux &          -8.967 &          -8.527 &          -7.940 &          -7.528 &          -7.450 &          -7.616 &          -7.966 &          -8.194 &          -8.381 &          -8.422 &          -8.609 \\
$1\sigma;\;(N_{stars})$ &       -9.803(5) &       -9.401(5) &       -8.650(7) &      \nodata(7) &       -8.772(7) &       -8.205(7) &       -8.092(7) &       -8.120(7) &       -8.214(7) &       -8.361(7) &       -8.623(6) \\

\cutinhead{M3--M4 Spectral Types}
\sidehead{\bf Stars}
             magnitudes &          18.466 &          16.957 &          14.924 &          12.889 &          11.966 &          11.653 &          11.418 &          11.322 &          11.275 &          11.345 &         \nodata \\
      median [log] flux &          -9.063 &          -8.616 &          -7.996 &          -7.558 &          -7.510 &          -7.684 &          -8.188 &          -8.442 &          -8.712 &          -9.153 &         \nodata \\
$1\sigma;\;(N_{stars})$ &      -10.046(3) &       -9.551(3) &      -8.764(12) &     \nodata(12) &      -8.797(12) &      -8.900(12) &      -9.343(12) &      -9.549(12) &      -9.771(12) &     -10.001(12) &      \nodata(0) \\
\sidehead{\bf Anemic Disks}
             magnitudes &          18.278 &          16.791 &          14.887 &          12.889 &          11.926 &          11.486 &          11.261 &          11.177 &          10.952 &          10.680 &           6.599 \\
      median [log] flux &          -8.987 &          -8.550 &          -7.982 &          -7.558 &          -7.494 &          -7.618 &          -8.126 &          -8.384 &          -8.582 &          -8.887 &          -8.664 \\
$1\sigma;\;(N_{stars})$ &       -9.699(3) &       -9.313(4) &       -8.791(5) &      \nodata(5) &       -8.705(5) &       -8.574(5) &       -8.676(5) &       -8.730(5) &       -8.784(5) &       -9.167(5) &      \nodata(1) \\
\sidehead{\bf Thick Disks}
             magnitudes &          18.209 &          16.888 &          14.964 &          12.889 &          11.991 &          11.489 &          10.809 &          10.452 &          10.021 &           9.123 &           6.342 \\
      median [log] flux &          -8.960 &          -8.589 &          -8.012 &          -7.558 &          -7.519 &          -7.619 &          -7.945 &          -8.094 &          -8.210 &          -8.264 &          -8.561 \\
$1\sigma;\;(N_{stars})$ &      \nodata(1) &      \nodata(1) &       -8.669(7) &      \nodata(7) &       -8.599(7) &       -8.338(7) &       -8.377(7) &       -8.473(7) &       -8.544(7) &       -8.675(7) &       -8.762(7) \\

\cutinhead{M4--M5 Spectral Types}
\sidehead{\bf Stars}
             magnitudes &          19.801 &          18.265 &          16.085 &          13.687 &          12.761 &          12.359 &          12.016 &          11.947 &          11.915 &          11.896 &         \nodata \\
      median [log] flux &          -9.597 &          -9.139 &          -8.461 &          -7.877 &          -7.827 &          -7.967 &          -8.428 &          -8.692 &          -8.967 &          -9.373 &         \nodata \\
$1\sigma;\;(N_{stars})$ &      -10.841(6) &       -9.853(9) &      -9.400(26) &     \nodata(27) &      -9.305(25) &      -9.457(25) &      -9.590(26) &      -9.812(26) &      -9.954(26) &     -10.258(25) &      \nodata(0) \\
\sidehead{\bf Anemic Disks}
             magnitudes &          19.966 &          18.392 &          16.136 &          13.687 &          12.781 &          12.380 &          12.037 &          11.943 &          11.861 &          11.553 &           8.566 \\
      median [log] flux &          -9.663 &          -9.190 &          -8.481 &          -7.877 &          -7.836 &          -7.975 &          -8.436 &          -8.690 &          -8.946 &          -9.236 &          -9.450 \\
$1\sigma;\;(N_{stars})$ &      -10.335(8) &      -9.795(11) &      -9.222(15) &     \nodata(16) &      -8.890(16) &      -8.833(16) &      -9.261(15) &      -9.522(15) &      -9.749(14) &      -9.921(12) &       -9.313(4) \\
\sidehead{\bf Thick Disks}
             magnitudes &          19.793 &          18.181 &          16.138 &          13.687 &          12.727 &          12.254 &          11.685 &          11.447 &          11.164 &          10.507 &           7.393 \\
      median [log] flux &          -9.593 &          -9.106 &          -8.482 &          -7.877 &          -7.814 &          -7.925 &          -8.295 &          -8.492 &          -8.667 &          -8.818 &          -8.981 \\
$1\sigma;\;(N_{stars})$ &      -10.426(8) &      -9.647(12) &      -9.093(18) &     \nodata(19) &      -9.016(19) &      -8.783(19) &      -9.174(19) &      -9.208(19) &      -9.338(18) &      -9.366(17) &      -9.309(14) \\

\cutinhead{M5--M6 Spectral Types}
\sidehead{\bf Stars}
             magnitudes &          21.230 &          19.424 &          16.873 &          14.345 &          13.529 &          13.115 &          12.718 &          12.619 &          12.603 &          12.886 &         \nodata \\
      median [log] flux &         -10.168 &          -9.603 &          -8.776 &          -8.140 &          -8.135 &          -8.269 &          -8.708 &          -8.961 &          -9.243 &          -9.769 &         \nodata \\
$1\sigma;\;(N_{stars})$ &      -10.588(7) &     -10.106(12) &      -9.573(16) &     \nodata(16) &      -9.391(16) &      -9.782(15) &      -9.695(15) &      -9.933(16) &     -10.159(16) &     -10.272(14) &      \nodata(0) \\
\sidehead{\bf Anemic Disks}
             magnitudes &          21.148 &          19.332 &          16.923 &          14.345 &          13.543 &          13.119 &          12.726 &          12.564 &          12.537 &          12.233 &           9.628 \\
      median [log] flux &         -10.136 &          -9.566 &          -8.796 &          -8.140 &          -8.141 &          -8.271 &          -8.711 &          -8.938 &          -9.216 &          -9.508 &          -9.875 \\
$1\sigma;\;(N_{stars})$ &      -10.791(8) &     -10.105(16) &      -9.650(25) &     \nodata(28) &      -9.549(26) &      -9.481(27) &      -9.523(28) &      -9.704(27) &      -9.946(25) &     -10.198(23) &      -10.172(2) \\
\sidehead{\bf Thick Disks}
             magnitudes &          21.042 &          19.171 &          16.852 &          14.345 &          13.521 &          13.064 &          12.579 &          12.276 &          11.923 &          11.123 &           7.830 \\
      median [log] flux &         -10.093 &          -9.502 &          -8.768 &          -8.140 &          -8.131 &          -8.249 &          -8.653 &          -8.823 &          -8.971 &          -9.064 &          -9.156 \\
$1\sigma;\;(N_{stars})$ &      -10.676(6) &     -10.238(11) &      -9.757(17) &     \nodata(19) &      -9.439(18) &      -9.580(18) &      -9.478(18) &      -9.528(18) &      -9.620(17) &      -9.647(17) &       -9.409(8) \\

\cutinhead{M6$+$ Spectral Types}
\sidehead{\bf Stars}
             magnitudes &          24.453 &          22.512 &          19.810 &          16.562 &          15.731 &          15.170 &          14.688 &          14.574 &          14.533 &          14.478 &         \nodata \\
      median [log] flux &         -11.458 &         -10.838 &          -9.951 &          -9.027 &          -9.016 &          -9.091 &          -9.496 &          -9.742 &         -10.015 &         -10.406 &         \nodata \\
$1\sigma;\;(N_{stars})$ &      -12.041(2) &      -11.551(5) &      -10.778(8) &      \nodata(8) &      -10.365(8) &      -10.508(8) &      -10.293(8) &      -10.594(8) &      -10.964(6) &      -12.237(3) &      \nodata(0) \\
\sidehead{\bf Anemic Disks}
             magnitudes &         \nodata &          22.497 &          19.621 &          16.562 &          15.667 &          15.120 &          14.592 &          14.429 &          14.060 &          13.977 &           9.669 \\
      median [log] flux &         \nodata &         -10.832 &          -9.875 &          -9.027 &          -8.990 &          -9.071 &          -9.458 &          -9.685 &          -9.826 &         -10.206 &          -9.892 \\
$1\sigma;\;(N_{stars})$ &      \nodata(0) &      -11.673(3) &     -10.531(15) &     \nodata(15) &     -10.153(15) &     -10.079(15) &     -10.292(14) &     -10.471(14) &      -10.349(7) &      -11.134(3) &      \nodata(1) \\
\sidehead{\bf Thick Disks}
             magnitudes &         \nodata &          22.504 &          19.765 &          16.562 &          15.710 &          15.159 &          14.390 &          13.982 &          13.493 &          12.724 &           9.965 \\
      median [log] flux &         \nodata &         -10.835 &          -9.933 &          -9.027 &          -9.007 &          -9.087 &          -9.377 &          -9.506 &          -9.599 &          -9.705 &         -10.010 \\
$1\sigma;\;(N_{stars})$ &      \nodata(0) &      -11.718(4) &     -10.543(11) &     \nodata(11) &     -10.359(10) &     -10.150(11) &      -9.916(11) &     -10.013(11) &      -10.089(8) &      -10.167(8) &      -10.964(2) \\
\enddata



\end{deluxetable}

\clearpage
\begin{deluxetable}{llclrc} 
\tablewidth{0pt}
\tablecaption{Passbands\label{tab:passbands} } 
\tabletypesize{}

\tablehead{
 \multicolumn{3}{c}{Data}  &
 \multicolumn{3}{c}{Flux Conversion}  \\
 \colhead{Passband}   &
 \colhead{System\tablenotemark{(a)}}     &
 \colhead{Source}     &
 \colhead{$\lambda_0\:(\micron)$}  &
 \colhead{\fnu (Jy)}  &
 \colhead{Ref.}         
}  

\startdata

 \Vl  & Landolt &  1,2,3     &  0.5423  & 3723.\phn\phn &   9    \\
 \Rc  & Cousins &  1,2,3     &  0.6410  & 3064.\phn\phn &  10    \\
 \Ic  & Cousins &  1,2,3     &  0.7890  & 2416.\phn\phn &  10    \\
 \Jtm & \tm     &  4,5,6,7   &  1.235   & 1594.\phn\phn &  11    \\
 \Htm & \tm     &  4,5,6,7   &  1.662   & 1024.\phn\phn &  11    \\
 \Ks  & \tm     &  4,5,6,7   &  2.159   & 666.8\phn &  11    \\
 \SIa & \irac   &  8         &  3.550   & 280.9\phn &  12    \\
 \SIb & \irac   &  8         &  4.493   & 179.5\phn &  12    \\
 \SIc & \irac   &  8         &  5.731   & 115.0\phn &  12    \\
 \SId & \irac   &  8         &  7.872   &  64.13    &  12    \\
 \SMa & \mips   &  8         &  23.68   &   7.3\phn &  13    \\

\enddata 

\tablenotetext{(a)}{Photometric system assumed for the conversion
from magnitude to flux. For example, all the near-infrared data
obtained from the FLAMINGOS instrument was zeropoint calibrated to
2MASS even though no formal color terms were included in that
magnitude calibration. Similarly, it is assumed that the Cousins
filter transmissions from 3 different surveys were identical.}
\tablerefs{:
 1:~\citet{1998ApJ...497..736H};~~	
 2:~\citet{1998ApJ...508..347L};~~	
 3:~\citet{1999ApJ...525..466L};~~	
 4:~\citet{2003yCat.2246....0C};~~	
 5:~\citet{2003AJ....125.2029M};~~	
 6:~\citet{2003ApJ...593.1093L};~~	
 7:~\citet{2005ApJ...618..810L};~~	
 8:~This paper;~~			
 9:~\citet{2003AJ....125.2645C};~~	
10:~\citet{1998A&A...333..231B};~~	
11:~\citet{2003AJ....126.1090C};~~	
12:~\citet{rea05};~~			
13:~SSC.
}

\end{deluxetable}

\end{document}